\documentclass[12pt,aps,epsf,twocolumns,numberedappendix,appendixfloats]{emulateapj}

\usepackage[breaklinks,colorlinks,urlcolor=blue,citecolor=blue,linkcolor=blue]{hyperref}
\usepackage{savesym}
\savesymbol{tablenum}
\usepackage{siunitx}
\restoresymbol{SIX}{tablenum}
\usepackage{mathtools}
\usepackage{physics}

\usepackage{graphicx}
\usepackage[FIGTOPCAP]{subfigure}
\usepackage{rotating}
\usepackage{epstopdf}
\usepackage{dcolumn}
\usepackage{bm}
\usepackage{float}
\usepackage{xcolor}

\newcommand\msun{\, \rm M_\odot}

\newcommand\kms{\, \rm km\,s^{-1}}

\newcommand\kpc{{\, \rm kpc}}
\newcommand\yr{{\, \rm yr}}

\newcommand{\eq}[1]{\begin{align}#1\end{align}}

\definecolor{purple}{rgb}{0.59, 0.44, 0.84}

\def\kms{\mathrm{km\,s}^{-1}}

\def\msun{\mathrm{M}_\odot}

\def\bsub{\begin{subequations}}
\def\esub{\end{subequations}}

\def\vmin{v_\mathrm{min}}

\def\barr{\begin{eqnarray}}
\def\earr{\end{eqnarray}}
\def\bsub{\begin{subequations}}
\def\esub{\end{subequations}}

\begin{document}

\title{Hypervelocity Stars from a Supermassive Black Hole-Intermediate-mass Black Hole binary}

\author{Alexander Rasskazov\altaffilmark{1}, Giacomo Fragione\altaffilmark{2}, Nathan W. C. Leigh\altaffilmark{3,4,5}, Hiromichi Tagawa\altaffilmark{1}, Alberto Sesana\altaffilmark{6}, Adrian Price-Whelan\altaffilmark{7}, Elena Maria Rossi\altaffilmark{8}}
 \affil{$^1$Institute of Physics, E\"{o}tv\"{o}s University, P\'azm\'{a}ny P. s. 1/A, Budapest, 1117, Hungary}
 \affil{$^2$Racah Institute for Physics, The Hebrew University, Jerusalem 91904, Israel} 
 \affil{$^3$Department of Physics and Astronomy, Stony Brook University, Stony Brook, NY 11794-3800, USA}
 \affil{$^4$Department of Astrophysics, American Museum of Natural History, New York, NY 10024, USA}
 \affil{$^5$Departamento de Astronom\'ia, Facultad de Ciencias F\'isicas y Matem\'aticas, Universidad de Concepci\'on, Concepci\'on, Chile}
\affil{$^6$School of Physics and Astronomy and Institute of Gravitational Wave Astronomy, University of Birmingham, Edgbaston B15 2TT, UK} 
\affil{$^7$Department of Astrophysical Sciences, Princeton University, 4 Ivy Lane, Princeton, NJ 08544, USA} 
\affil{$^8$Leiden Observatory, Leiden University, PO Box 9513, NL-2300 RA Leiden, the Netherlands} 

 \begin{abstract}
In this paper we consider a scenario where the currently observed hypervelocity stars in our Galaxy have been ejected from the Galactic center as a result of dynamical interactions with an intermediate-mass black hole (IMBH) orbiting the central supermassive black hole (SMBH). By performing 3-body scattering experiments, we calculate the distribution of the ejected stars' velocities given various parameters of the IMBH-SMBH binary: IMBH mass, semimajor axis and eccentricity. We also calculate the rates of change of the BH binary orbital elements due to those stellar ejections. One of our new findings is that the ejection rate depends (although mildly) on the rotation of the stellar nucleus (its total angular momentum). We also compare the ejection velocity distribution with that produced by the Hills mechanism (stellar binary disruption) and find that the latter produces faster stars on average. Also, the IMBH mechanism produces an ejection velocity distribution which is flattened towards the BH binary plane while the Hills mechanism produces a spherically symmetric one. The results of this paper will allow us in the future to model the ejection of stars by an evolving BH binary and compare both models with \textit{Gaia} observations, for a wide variety of environments (galactic nuclei, globular clusters, the Large Magellanic Clouds, etc.). 
 \end{abstract}

\section{\label{section:introduction} Introduction}

In the last decade, stars with extreme radial velocities have been discovered in the Galactic halo, the so-called hypervelocity stars (HVSs). They were first predicted by \citet{hills88} as a natural consequence of the presence of a super massive black hole (SMBH) in the Galactic Centre (GC) that tidally disrupts binary stars approaching too close. However, they were not detected until 2005, when the first HVS was observed by \citet{brw05} escaping the Milky Way (MW) with a heliocentric radial velocity of $\sim 700\kms$. \citet{GPZS2005} estimated an ejection velocity of $\gtrsim 1000\kms$ from the GC.  Since then, $\sim 20$ HVSs have been detected by the Multiple Mirror Telescope Survey out to a Galactocentric distance of $\sim 120\kpc$ and with velocities of up to $\sim 700\kms$ \citep{brw06,brw14}.  With an estimated ejection rate of $\sim 10^{-6}-10^{-4}\yr^{-1}$\citep{Yu2003,per09,zhang2013}, HVSs remain rare objects.

{The European mission \textit{Gaia}\footnote{http://sci.esa.int/gaia/} has recently revolutionized astrometry and promised to discover new HVSs \citep{brw15,mar17}.  The recent second data release (DR2) has provided positions, parallaxes and proper motions for more than $\sim 1$ billion stars, along with radial velocities for $\sim 7$ million stars \citep{gaia18}, thus offering an unprecedented opportunity to study the Galactic population of HVSs. Exploiting these data, \citet{brown18} showed that only the fastest B-type HVSs (radial velocities $\gtrsim 450\kms$) have orbits surely consistent with a GC origin, while other HVSs have ambiguous origins. Gaia new astrometry also allowed to reject most of the -- already highly debated -- late type HVS candidates \citep{bou18} and add evidence on the Large Magellanic origin of HVS3 \citep{erk18}. As expected, discovery of new unbound HVSs has not yet happened because the sample of stars in DR2 with radial velocity is relatively too small and nearby for the rarity of HVSs and their expected distance distribution \citep{march18}. On the other hand, bound HVSs may have been found, although their metallicity and age make them hard to set apart from the halo star population \citep{hat18}. The real treasure trove should be found in the rest of the catalogue with 5 dimension parameter space information, that requires more sophisticated data mining methods in order to select a manageable number of candidates to be spectroscopically followed up \citep{mar17,brom18}.}

{Theoretically}, many properties of the observed (and expected) HVSs remain poorly understood, including the dominant ejection mechanism. Several different mechanisms have been proposed to produce HVSs, besides the standard Hills binary disruption mechanism, still regarded as the favoured model. The main alternative was first proposed and discussed by \citet{Yu2003}, who suggested the ejection of HVSs as a consequence of star slingshots involving a massive black hole binary composed of the $4\times 10^6\msun$ SMBH already observed in the GC (i.e. SgrA$^*$) and a putative secondary SMBH or intermediate mass black hole \citep[IMBH;][]{baumg2006,levin2006,fragk18,fglk18}, possibly delivered by infalling globular clusters due to dynamical friction in their host galaxy \citep{arca2018,fragk18,fglk18}. A scenario subsequently explored in a series of papers by Sesana and collaborators, considering either the ejection of unbound stars \citep{Sesana2006}, or the erosion of a pre-existent bound stellar cusp \citep{2008ApJ...686..432S}. Encounters with nearby galaxies \citep{GPZ2007,bou2017}, supernova explosions \citep{zub2013,tau2015}, interactions of star clusters with single or binary SMBHs or IMBHs \citep{cap15,fra16,fck17}, {nuclear spiral arms perturbation} \citep{hamp2017}, and the dynamical evolution of a disk orbiting the SMBH in the GC \citep{sub16} could all produce HVSs.

In this paper, we perform scattering experiments to determine the HVS ejection rate and also the distribution of their ejection velocities for both the Hills and IMBH binary companion scenarios. For the SMBH-IMBH binary scenario, we also calculate the evolution rate of its semimajor axis and eccentricity. Compared to previous works, we extend the scattering experiment parameter space to binary mass ratios as small as $10^{-4}$, and also consider the cases of corotating and counterrotating stellar nuclei. We correct an error in the calculations of the eccentricity evolution rate by \citet{Sesana2006}. {Another improvement is that for our scattering experiments we use the \textsc{ARCHAIN} algorithm. \textsc{ARCHAIN} is a fully regularized scheme that adopts a transformation in the Hamiltonian of the system and is able to model the evolution of binaries of arbitrary mass ratios and eccentricities with high accuracy over long periods of time \citep{mik06,Mikkola2008}.}

The paper is organized as follows. In Section \ref{section:mechanisms} we discuss various phenomena which could generate HVSs. In subsequent sections we numerically simulate two of those mechanisms: slingshot ejections by an SMBH-IMBH binary (Sections \ref{section:methods} and \ref{section:results}) and the Hills mechanism (Section \ref{section:hills}). We conclude and discuss future work in Section \ref{section:conclusions}.

\section{Mechanisms for producing hypervelocity stars}
\label{section:mechanisms}

In this section, we review the different mechanisms that could form hypervelocity stars in our Galaxy, along with their predicted observational signatures.  The predicted observational properties expected for each mechanism are summarized in Table~\ref{tab:summary}. 
\begin{table*}
\caption{Summary of predicted observational properties for each HVS mechanism.}
\centering
\begin{tabular}{lcccc}
\hline
Mechanism & Origin & Object type & HVS binaries\\
\hline\hline
Hills mechanism & GC, LMC, globular clusters & Young, old & No \\
SMBH-IMBH	& GC, LMC & Young, old & Yes \\
IMBH-BH & Globular clusters, LMC &  Young, old & Unlikely \\
SN explosions & Galactic disk, Globular clusters & WD & No \\
\hline
\end{tabular}
\label{tab:summary}
\end{table*}

\subsection{The Galactic Centre}

The first mechanism proposed to produce hypervelocity stars is called the Hills mechanism \citep{hills88}, and involves the disruption of binary star systems by a central super-massive black hole.  As already discussed, a binary is required in order to provide a reservoir of negative orbital energy by leaving one of the binary components bound to the SMBH, such that a large positive kinetic energy for the ejected star is allowed by energy conservation.  This mechanism can produce hypervelocity stars with velocities $\gtrsim \SI{1000}{km/s}$, depending on the SMBH mass and the properties of the stars (see Section \ref{section:hills}).

This mechanism predicts an isotropic distribution of HVSs with 3-D velocities pointing back to the GC, assuming the initial binary stars are injected into the SMBH loss cone isotropically from a spherical nuclear star cluster.   If the binaries are injected from a stellar disk, then this should be reflected accordingly in the final velocity distribution of observed hypervelocity stars \citep{2013ApJ...768..153Z,sub16}.

The Hills mechanism assumes an isolated SMBH, but it remains possible that an IMBH is also present in the GC and orbits Sgr A*.  \citet{Yu2003} constrained the allowed orbital properties of such an hypothesized IMBH \citep[see also][]{2009ApJ...705..361G}.  Interestingly, if an IMBH exists, this opens up another channel for producing HVSs.
Specifically, if single stars pass close to the SMBH-IMBH pair, they can be flung out of the Galaxy at roughly the orbital speed of the IMBH (i.e., up to thousands of km s$^{-1}$), draining orbital energy from the BH binary. 

This second mechanism for producing HVSs also predicts HVSs with 3-D velocities that point back to the GC.  But, assuming isotropic injection of single stars in to the SMBH-IMBH loss cone, the predicted velocity distribution should be skewed such that it is preferentially aligned with the orbital plane of the SMBH-IMBH binary. As shown in this paper using sophisticated $3$-body integrations, the characteristic final ejection velocities associated with this mechanism should be lower than predicted by the standard Hills mechanism. {Note, however, that while we compute the ejected velocities at fixed binary separations, to compute the overall expected distribution the IMBH orbit has to be self-consistently evolved \cite[][]{2018MNRAS.476.4697M,2018arXiv181005312D}. Moreover, an inspiralling IMBH will also disrupt the cusp of stars bound to SgrA$^*$, a process that can lead to a burst of stellar ejections at very high velocities \citep{2008ApJ...686..432S}}, which we don't include in our model.  

Finally, this is the only mechanism that can generate HVS binaries \citep{2007ApJ...666L..89L,2009MNRAS.392L..31S,wang18}, offering a potential means of constraining the possible presence of an IMBH in the Galactic Center.

\subsection{Globular clusters}

In Galactic globular clusters, the same mechanisms for producing HVSs in the GC could also be operating, provided IMBHs are present.  The Hills mechanism could operate, if binary stars drift within the loss cone of any existing IMBHs \citep*{fgu18,subr2019} {or if single stars interact with a binary IMBH}. As found in \citet{leigh14}, if stellar-mass BHs are also present in globular clusters hosting IMBHs, then the most massive BH remaining in the system will quickly end up bound to the IMBH in a roughly Keplerian orbit but with a very high orbital eccentricity.  This IMBH-BH binary could then produce HVSs by interacting directly with single stars in the cores of globular clusters, in close analogy with the production of HVSs by an SMBH-IMBH binary in the GC. { Finally, the few-body dynamical interaction of stars could also eject stars with a typical velocity of $\sim$ tens $\kms$ \citep{persubr2012,ohkroupa2016}. }

Ejection of HVSs from globular clusters predicts a mean HVS velocity that is much smaller than predicted for HVSs with a GC origin, due to the much smaller masses of IMBHs compared to SMBHs.  It also predicts HVS velocities distributed roughly isotropically on the sky and pointing inward toward the central parts of the Galaxy and its disc, provided every globular cluster in the Galaxy contributes commensurably to HVS production (and assuming an isotropic distribution of globular clusters in the MW halo).  

Ejections from this environment also predict the widest distribution of HVS velocities, since MW globular clusters have orbital velocities of order $\sim 200$ km s$^{-1}$ but with a wide range of orbital motions through the Galaxy.  Hence, the predicted distribution of HVS velocities must be broadened accordingly by the mean globular cluster orbital velocity.  

\subsection{The Large Magellanic Clouds}

Whether or not the Large Magellanic Clouds (LMC) are home to one or more massive BHs is an actively debated topic \citep{bou16,bou2017}.  But, if the LMC is home to any massive BHs, then we might expect HVSs produced analogously to those discussed above, only with 3-D velocities pointing back to the LMC instead of the Milky Way. {As mentioned before, two such candidate HVSs have been identified \citep{erk18,hat18}.} 

The orbital velocity of the LMC about the MW is of order $\sim$ 300 km s$^{-1}$ \citep{kal13}.  Hence, this mechanism for HVS production also predicts a mean HVS velocity shifted to higher velocities by of order $\sim$ 300 km s$^{-1}$ relative to the Galactic rest frame.  Note that this distribution is simply shifted to higher velocities, and does not suffer from the same broadening described above for HVSs produced from Galactic globular clusters. Thus, we might naively expect a comparable maximum velocity for both the globular cluster and LMC ejection scenarios, but a much broader velocity distribution for the former mechanism.

Interestingly, if there is an IMBH/SMBH in the LMC and it has a close binary BH companion, then (after correcting for the orbital motion of the LMC relative to the Milky Way, and the MW's gravitational influence on the distribution of ejection velocities) the resulting distribution of HVSs may either resemble a torus for circular orbits (i.e., ejected preferentially in the orbital plane of the binary but with some dispersion) or a one-sided jet for eccentric orbits \citep[e.g.][]{Quinlan1996,Sesana2006}.  
If, on the other hand, an IMBH/SMBH is present in the LMC but has no binary companion, then the expected distribution of HVSs should be isotropic relative to the position of the IMBH/SMBH in the LMC. 

\subsection{Kicks during supernovae explosions in binaries}

HVSs can also be produced in binary star systems when one of the companions explodes as a supernova (SN), ablating the stellar progenitor completely.  This blast wave should quickly flow beyond the original orbit of the companion, such that is escapes to infinity with a final velocity of order the Keperian velocity at the time of explosion (in the direction of motion the exploding WD).  However, in order for large velocities of order $\gtrsim$ 1000 km s$^{-1}$ to be achieved, both binary companions must be compact objects prior to the supernova explosion.  Hence, most of the HVSs are expected to be white dwarfs (WDs) originating from compact accreting WD-WD binaries \citep{shen18}.

This mechanism predicts HVSs with 3-D velocities originating from the disk of our Galaxy, since this is where the majority of the progenitor WD-WD binaries are expected to reside.  Most of these should be born from more massive stars with shorter lifetimes  and hence associated with young star-forming regions, since the most massive WDs are needed in order to accrete enough material from their binary companions to exceed the Chandrasekhar limit.  Recently, three candidate HVS WDs were identified by \citet{shen18} using GAIA data, which the authors postulate to have formed from kicks during SN explosions in WD-WD binaries. 

\section{Scattering experiments: method}
\label{section:methods} 

To calculate the ejected star velocities we use scattering experiments analyzed using the method originally presented in \cite{Quinlan1996} and later built upon by \cite{Sesana2006} and \cite{Rasskazov2017}. 
We assume every star is initially unbound from the SMBH-IMBH binary and approaches it from infinity. Then we follow its interaction with the binary until the star reaches the local escape speed, defined such that the star has positive total energy at spatial infinity. At the beginning of every simulation, we specify the following: 

\begin{enumerate}
\item Binary mass ratio $q = M_2/M_1 < 1$.
\item Binary eccentricity $e$.
\item Initial stellar velocity (at infinity) $v$.
\item Stellar impact parameter $p$.
\item Two angles defining the initial orbital plane of the star $\{\theta, \phi\}$.
\item One angle defining the direction of the initial stellar velocity with respect to its orbital plane $\psi$.
\item Initial orbital phase of the binary $\psi_b$.
\end{enumerate}

The stellar orbit integrations are carried out using ARCHAIN, an implementation of algorithmic regularization developed specifically to treat small-$N$ systems \citep{Mikkola2008}. The incoming star is considered massless which further simplifies the problem: the binary's center of mass is always at the center of the coordinate system. 

To speed up the computations, the stellar orbit is treated as Keplerian whenever the star is farther than 50 binary semimajor axes from the center of mass; whenever it leaves that sphere, we 1) calculate its coordinates and the time at which it re-enters the sphere, 2) calculate the binary phase at that time, and then 3) restart the 3-body integration with the new initial conditions. The reduction in the required CPU time can reach an order of magnitude for the runs with $q\ll1$ when a significant fraction of the stars get captured on to very loosely bound orbits (semimajor axes $>100a$) and perform hundreds of revolutions before finally being ejected. We confirm that this approximation does not introduce any noticeable bias into the distribution of HVS parameters. 

During a simulation, we use the binary semimajor axis $a$ as the unit of distance and the orbital velocity $v_0=\sqrt{GM/a}$ (where $M=M_1+M_2$ is the total binary mass) as the unit of velocity, with the binary period being set equal to $2\pi$. For this reason, our simulation results do not depend on the binary mass and semimajor axis as long as $p/a$ and $v/v_0$ are fixed. 

It has been previously shown that a binary only becomes efficient at ejecting stars at semimajor axes $a<a_h=GM_2/(4\sigma^2)$. For that reason, following \citet{Sesana2006}, for every pair of $\{q,e\}$ we sample $v$ in the range $\num{3e-3}\sqrt{q/(1+q)}<v/v_0<30\sqrt{q/(1+q)}$ with 80 logarithmically sampled points. This allows us to sample a Maxwellian distribution (using rejection sampling) for a relevant range of binary hardness at any $q$:
$a/a_h \sim 4(v/v_0)^2(1+q)/q \sim 10^{-5} \dots  10^3$. For every value of $v$ we perform $\num{5e4}$ scattering simulations, where $\cos\theta$ and all the other angles are uniformly distributed, and $p^2$ is uniformly distributed in the range $[0,25(1+0.4v^{-2})a^2]$ which corresponds to a pericenter distance range $[0,5a]$. In this way, for every $\{q,e\}$, we perform a total of $\num{4e6}$ simulations, { which took about a week on a desktop computer}. 

The simulations are stopped when one of the following conditions is met:
\begin{enumerate}
\item The star leaves the sphere of radius $50a$ with positive energy.
\item The total interaction timescale exceeds $10^{10}\,\si{yr}$ (assuming parameters similar to the Milky Way $M=\SI{4e6}{\msun}$, $\sigma = \SI{70}{km/s}$ and $a=a_h$).
\item The total time the star spends inside the sphere of radius $50a$ (when its orbit is not treated as Keplerian) exceeds $10^5$ time units (\num{1.6e4} binary orbital periods). 
\end{enumerate}

Only in the first case the star is treated as ejected. Its velocity is calculated at infinity with its orbit being extrapolated from the radius $50a$ as Keplerian (hyperbolic). Stars falling into the two latter categories are discarded and not included in the simulated HVS dataset. Their fraction is usually the highest for the simulations with low $v$, low $q$ and high $e$; for the fastest velocity bin it is always zero (the stars simply fly by).  
In the slowest velocity bin, 0.5\%-3\% and 6\%-9\% of all simulations fall into categories 2 and 3, respectively.

Given the scattering experiment data described above, it is possible to introduce the rotation of the stellar nucleus in the same way as in \cite{Rasskazov2017} or \cite{Gualandris2012}:
\begin{enumerate}
\item Choose a direction for the stellar rotation axis.
\item Divide all the stars into  ``corotating'' and ``counterrotating'' depending on the sign of the projection of their initial angular momentum on to the axis of rotation.
\item Choose the fraction of corotating stars $\eta\in[0,1]$, and remove some stars accordingly.
\end{enumerate}
In what follows, we check the dependence of the hypervelocity star parameters on $\eta$ as well as on the mutual orientation of the binary and star orbital planes.

\bgroup
\def\arraystretch{1.5}
\begin{table}
\caption{Best-fit parameters for the hardening rate $H$ (Eq.~\ref{eq:H})} 
\centering
\begin{tabular}{c|cccc}
\hline\hline
& rotation & $A_H$ & $a_{0,H}/a_h$ & $\gamma_H$ \\ \hline
& counter- & 11.1 & 2.08 & $-0.72$\\
$q=10^{-4},\, e=0$ & none & 16.3 & 3.34 & $-0.78$\\ 
& co- & 21.6 & 4.34 & $-0.82$\\ \hline
$q=10^{-4},\, e=0.9$ & none & 18.0 & 2.46 & $-0.77$\\ \hline
& counter- & 11.5 & 2.58 & $-0.78$\\ 
$q=10^{-3},\, e=0$ & none & 16.8 & 3.21 & $-0.73$\\ 
& co- & 22.3 & 3.74 & $-0.73$\\ \hline
$q=10^{-3},\, e=0.9$ & none & 18.3 & 2.43 & $-0.75$\\ \hline
\end{tabular}
\label{table:H}
\end{table}
\egroup

\bgroup
\def\arraystretch{1.5}
\begin{table*}
\caption{Best-fit parameters for the mass ejection rate $J$ (Eq.~\ref{eq:J})} 
\centering
\begin{tabular}{c|cccccc}
\hline\hline
& rotation & $A_J$ & $a_{0,J}/a_h$ & $\alpha_J$ & $\beta_J$ & $\gamma_J$ \\ \hline
& counter- & 0.265 & 0.188 & $-0.353$ & 5.463 & $-0.237$\\
$q=10^{-4},\, e=0$ & none & 0.226 & 0.221 & $-0.300$ & 3.540 & $-0.355$\\ 
& co- & 0.211 & 0.245 & $-0.256$ & 2.788 & $-0.463$\\ \hline
$q=10^{-4},\, e=0.9$ & none & 0.852 & 0.180 & 0 & 0.703 & $-2.433$\\ \hline
& counter- & 0.258 & 0.181 & $-0.353$ & 5.042 & $-0.245$\\ 
$q=10^{-3},\, e=0$ & none & 0.218 & 0.209 & $-0.304$ & 3.948 & $-0.300$\\ 
& co- & 0.199 & 0.231 & $-0.268$ & 3.465 & $-0.346$\\ \hline
$q=10^{-3},\, e=0.9$ & none & 0.693 & 0.165 & $-0.034$ & 0.773 & $-2.064$\\ \hline
\end{tabular}
\label{table:J}
\end{table*}
\egroup

\begin{figure*}
	\centering
	\subfigure{\includegraphics[width=0.49\textwidth]{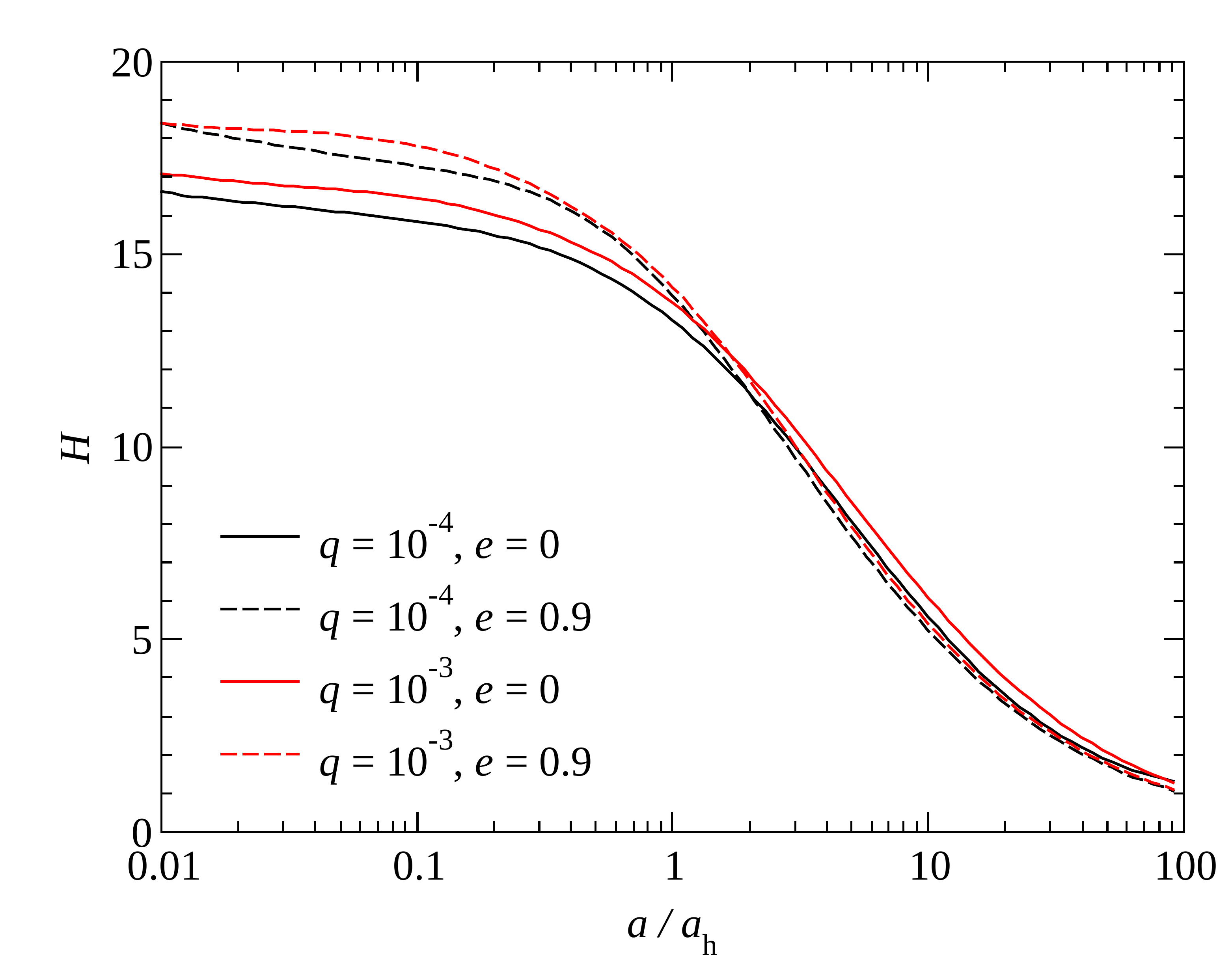}}
	\subfigure{\includegraphics[width=0.49\textwidth]{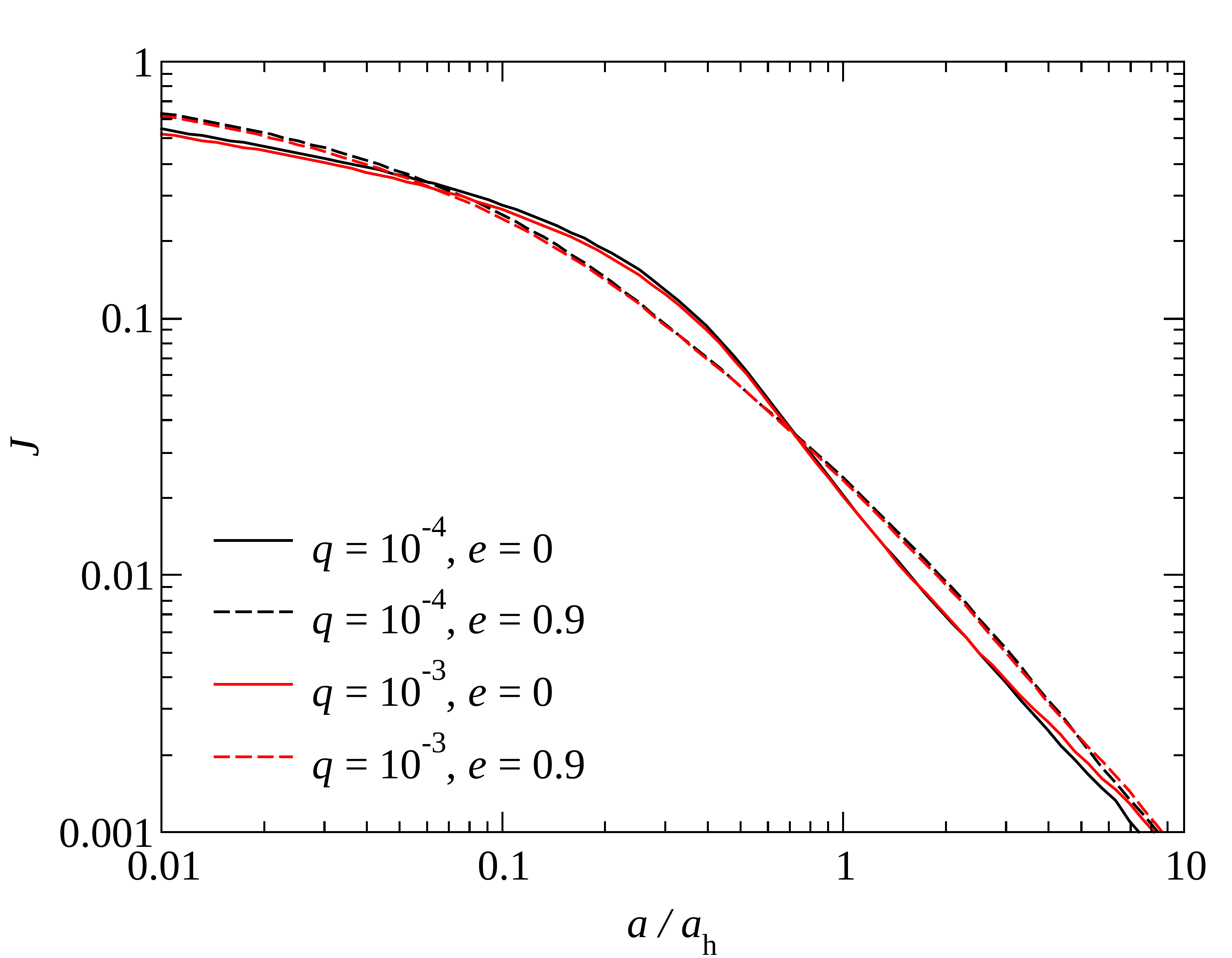}}
	\subfigure{\includegraphics[width=0.49\textwidth]{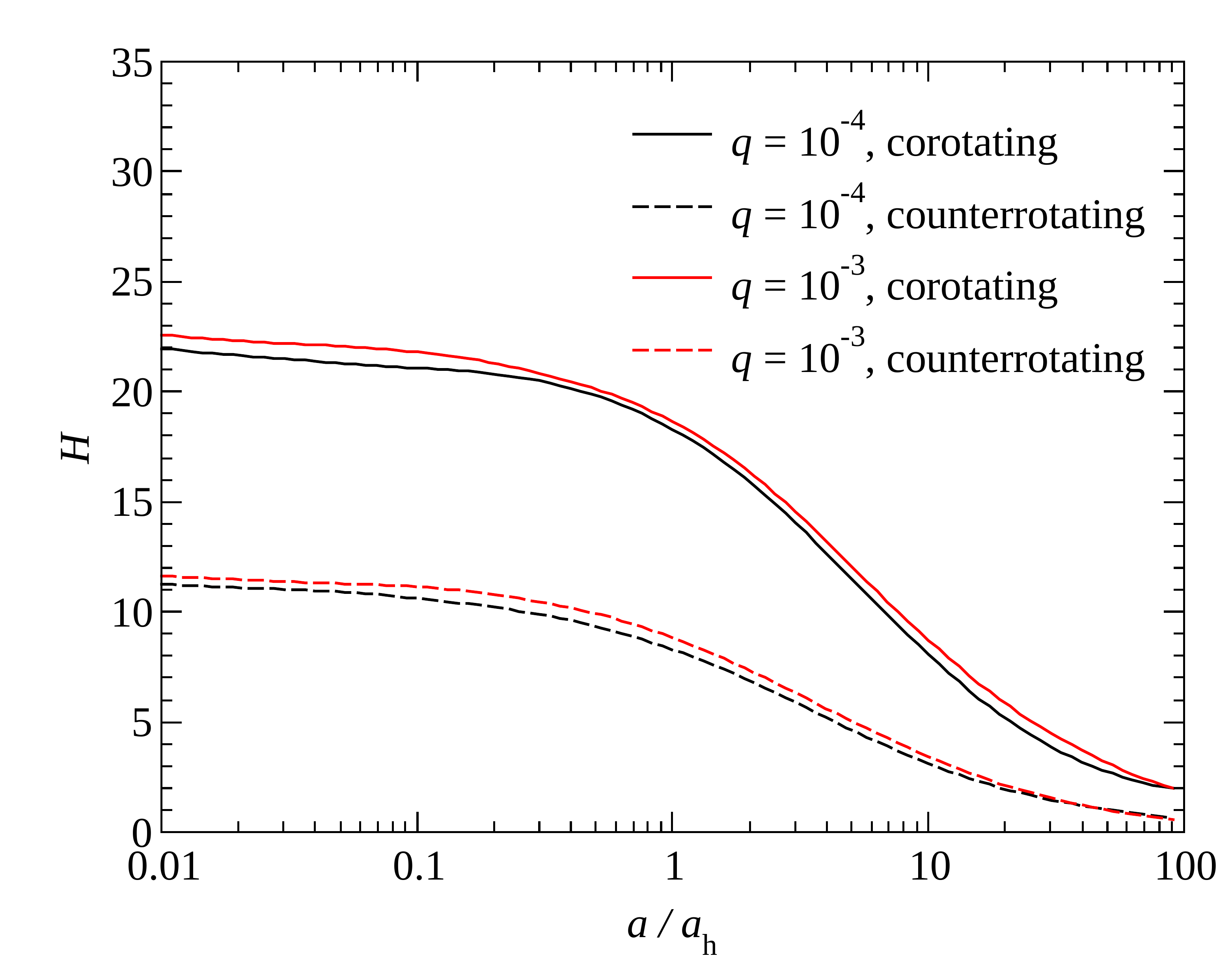}}
	\subfigure{\includegraphics[width=0.49\textwidth]{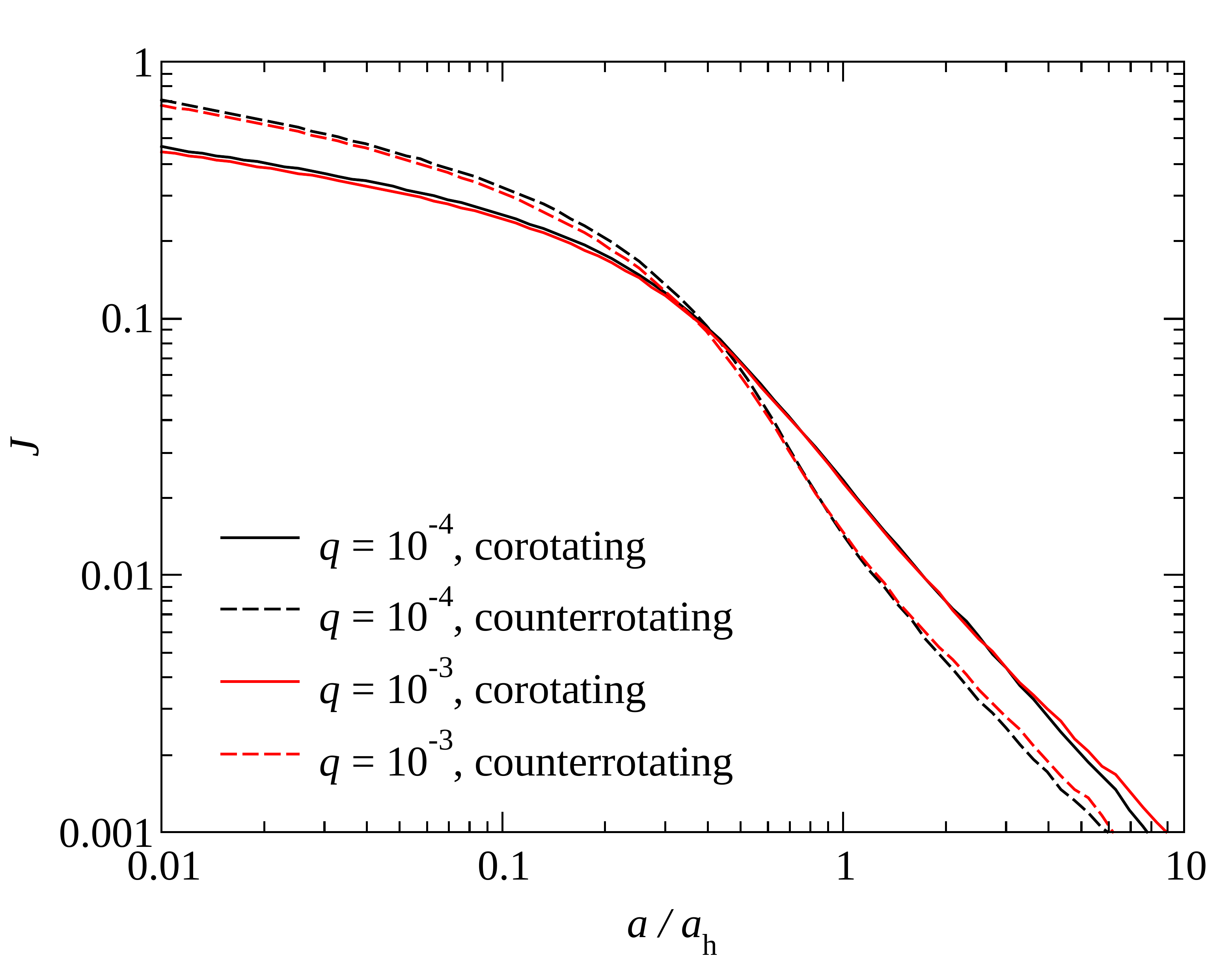}}
\caption{
The dependence of the hardening and ejection rates on the binary hardness for a non-rotating stellar nucleus with different values of $q$ and $e$ (top) and $e=0$ with different assumptions about the rotation of the host nuclear star cluster (bottom).
} 
\label{fig:HJ}
\end{figure*}

\section{Scattering experiments: results}
\label{section:results}

\begin{figure*}
	\centering
	\subfigure{\includegraphics[width=0.49\textwidth]{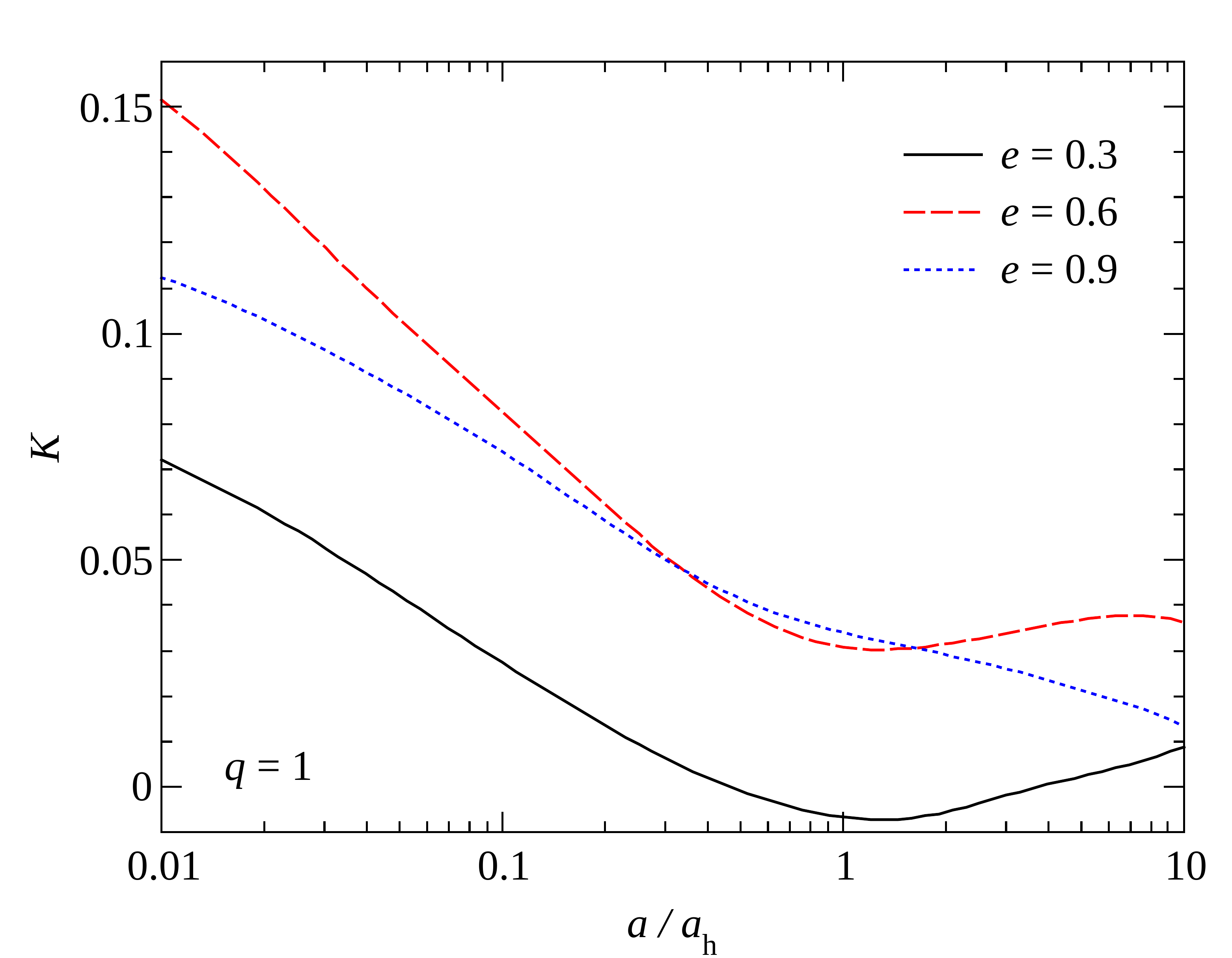}}
	\subfigure{\includegraphics[width=0.49\textwidth]{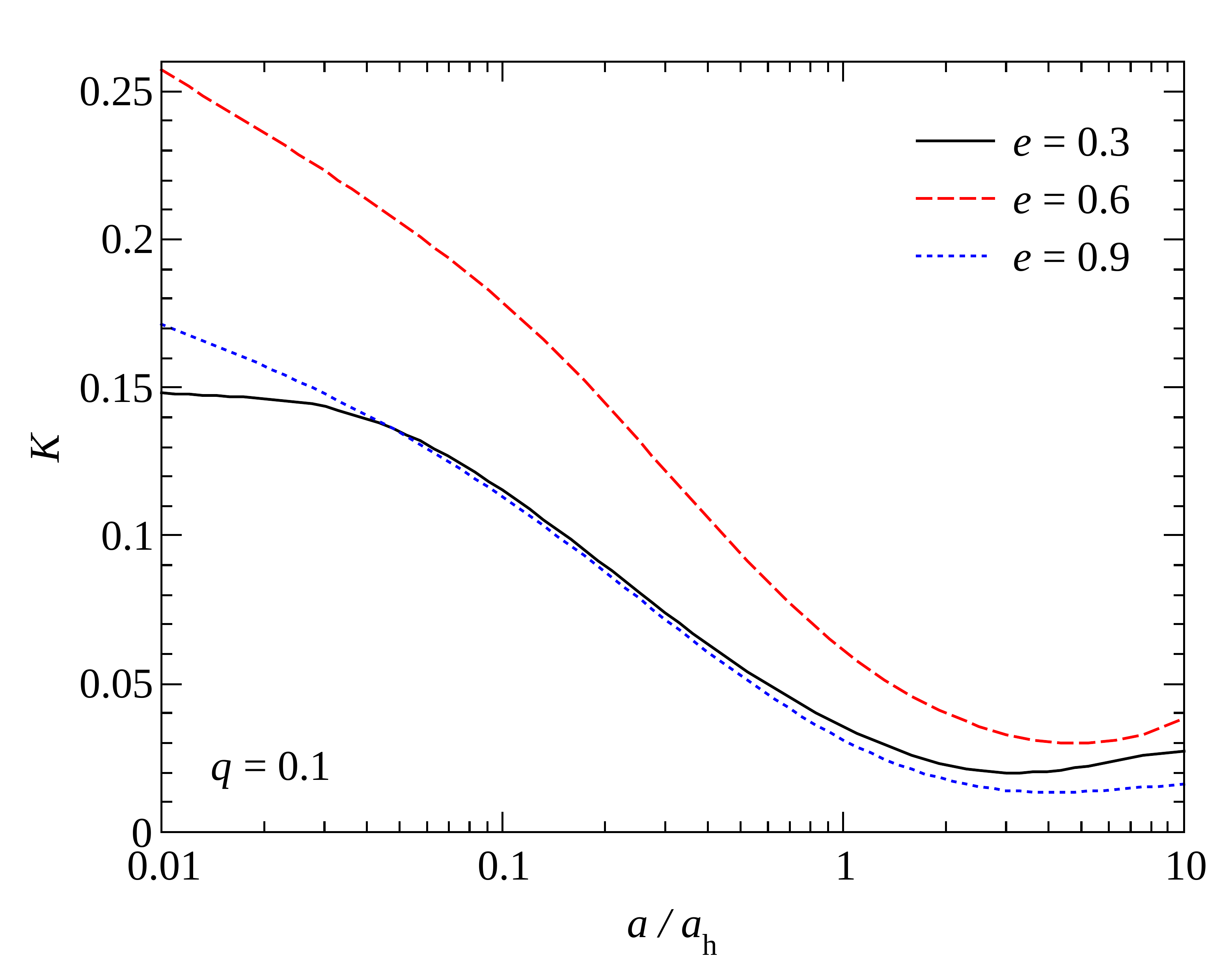}}
	\subfigure{\includegraphics[width=0.49\textwidth]{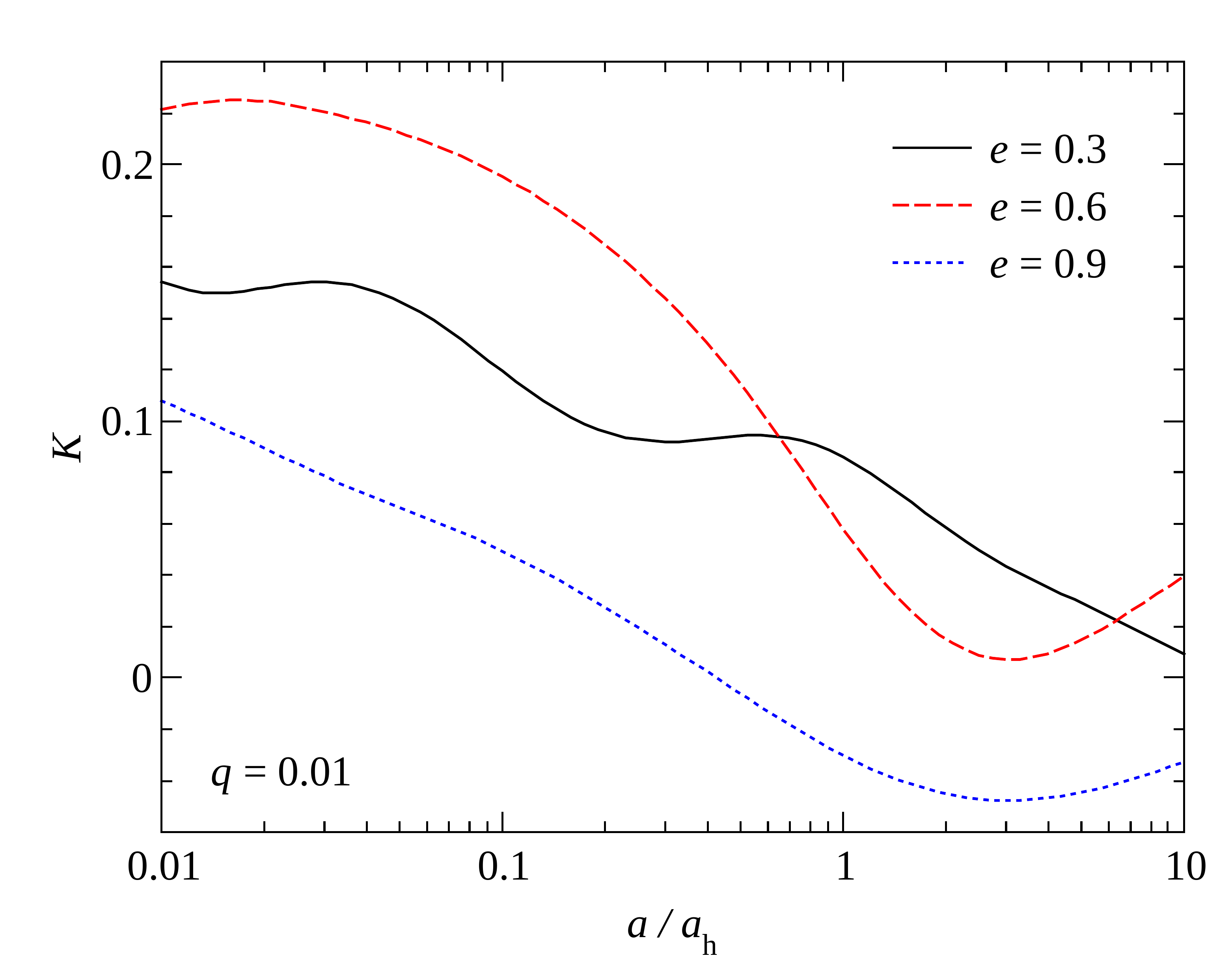}}
	\subfigure{\includegraphics[width=0.49\textwidth]{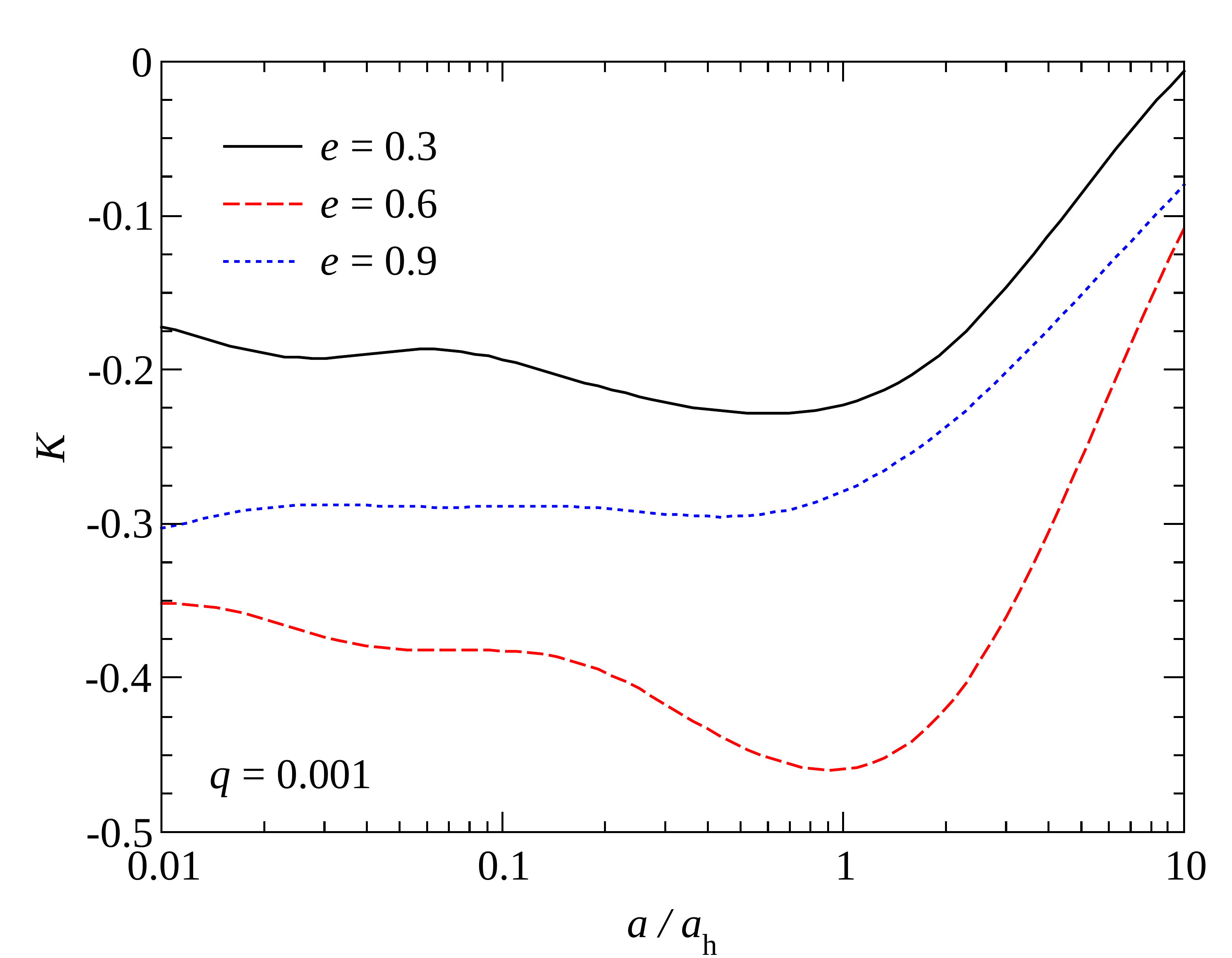}}
	\subfigure{\includegraphics[width=0.49\textwidth]{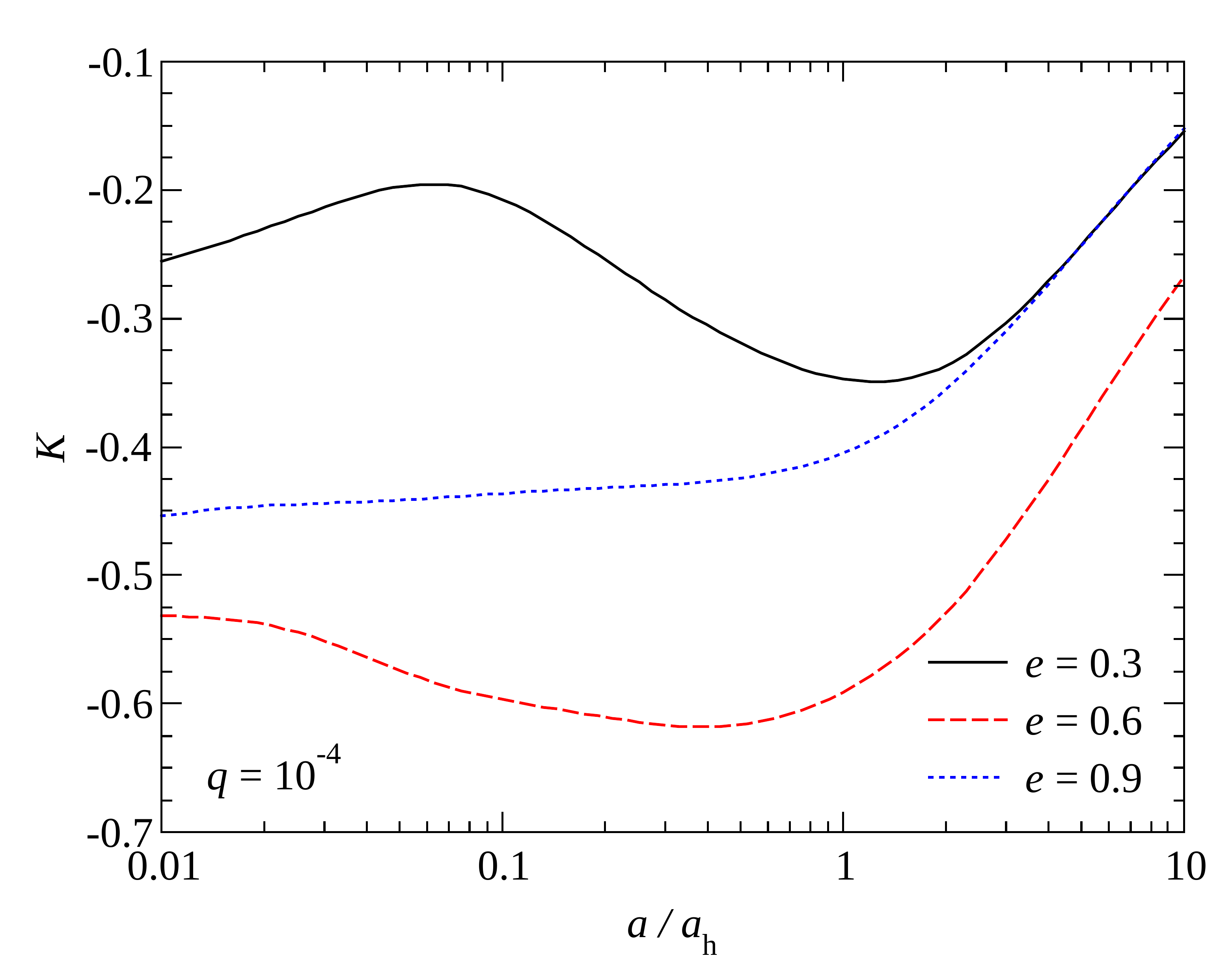}}
\caption{
The dependence of the rate of eccentricity evolution on the binary hardness for a non-rotating stellar nucleus with different values of $q$ and $e$.
}
\label{fig:K}
\end{figure*}

\begin{figure}
	\centering
\subfigure{\includegraphics[width=0.49\textwidth]{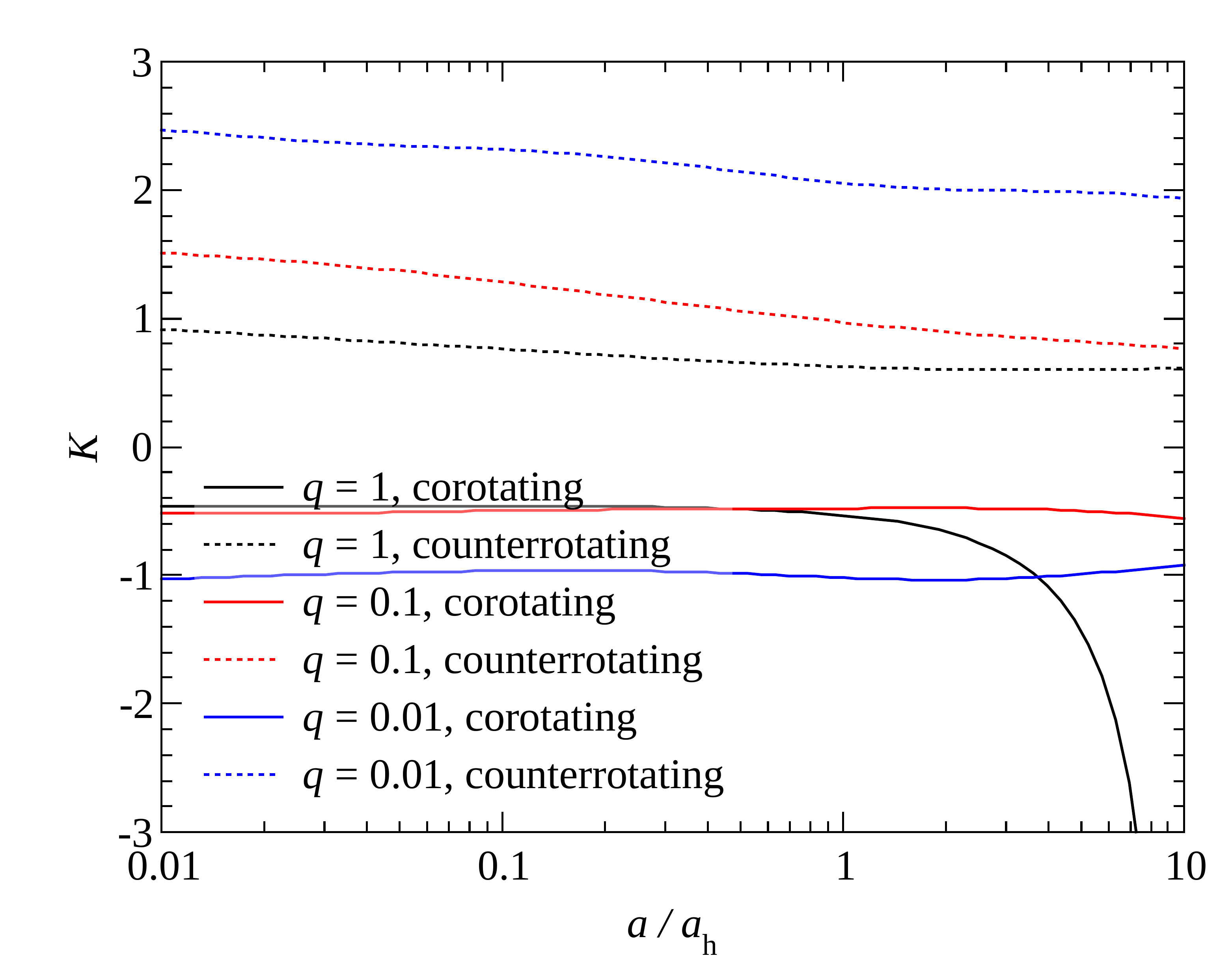}}
\subfigure{\includegraphics[width=0.49\textwidth]{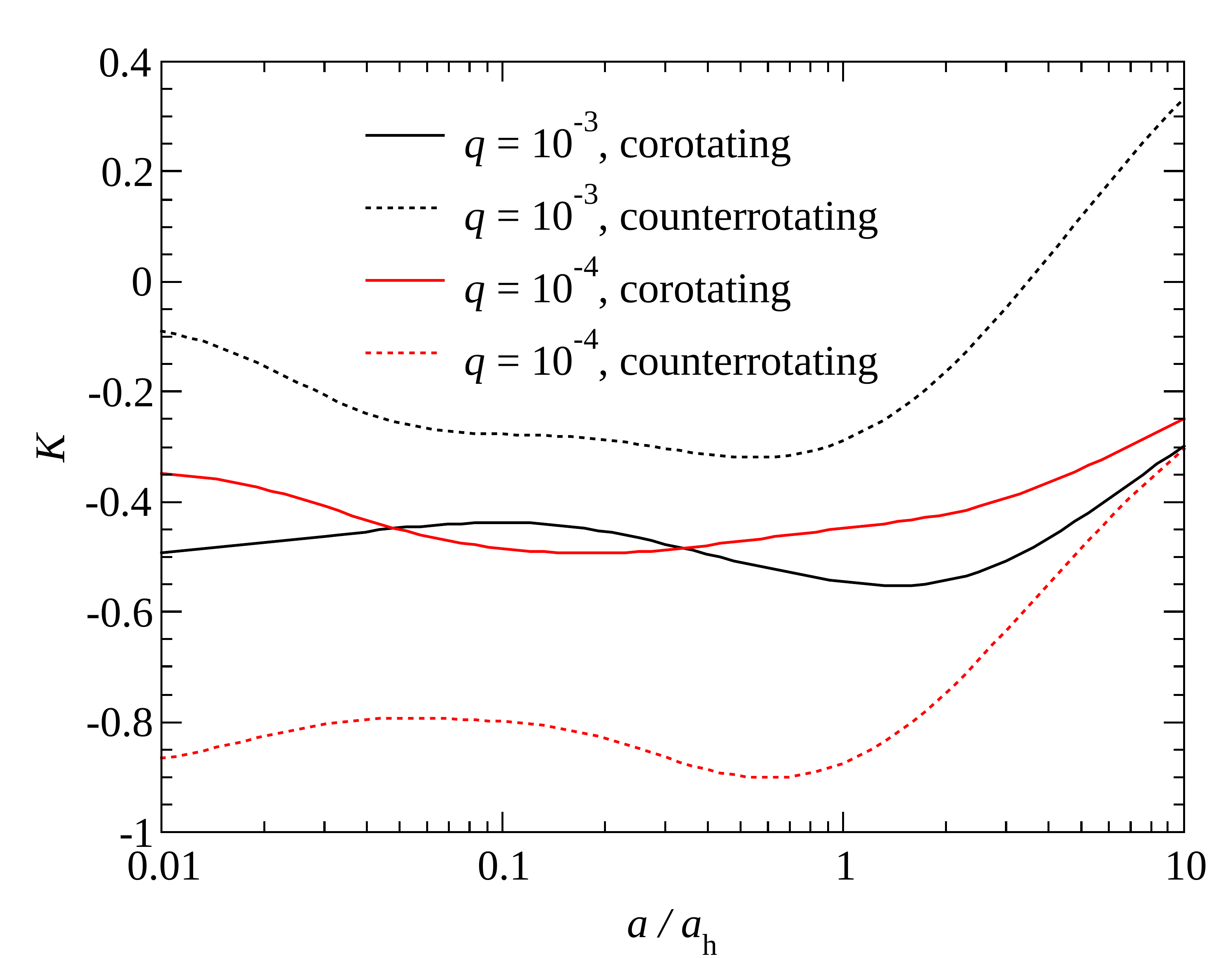}}
\caption{
The dependence of the rate of eccentricity evolution on the binary hardness for $e=0.6$ and various assumptions about the rotation of the host cluster.
}
\label{fig:K1}
\end{figure}

In this section we calculate various dimensionless parameters of the binary evolution necessary to model the distribution of HVS velocities. In accord with previous works {\citep{Quinlan1996}}, we define the hardening rate $H$, mass ejection rate $J$ and the rate of eccentricity evolution $K$ as
\bsub
\barr
H &=& \frac{\sigma}{G\rho}\dv{}{t}\qty(\frac{1}{a}),\\
J &=& \frac{1}{M} \dv{M_\mathrm{ej}}{\ln(1/a)},\\
K &=& \dv{e}{\ln(1/a)} \label{eq:K-definition}
\earr
\esub
where $M_\mathrm{ej}$ is the stellar mass ejected by the binary{, where we consider a star ``ejected'' if its final velocity is above $5.5\sigma$ (approximately the escape velocity from the MW center). All of these quantities only depend on dimensionless parameters such as $a/a_h$ but not on $M$ or $a$. Knowing $H$, $J$ and $K$, we can calculate the change rates of $a$ and $e$ as well as the ejection rate in the following way:
\eq{
\dv{a}{t} &= -\frac{a}{t_h},\\
\dv{M_\mathrm{ej}}{t} &= \frac{JM}{t_h},\\
\dv{e}{t} &= \frac{K}{t_h},\\
t_h &\equiv \frac{\sigma}{HG\rho a} \approx \SI{13}{Myr} \frac{\sigma}{\SI{100}{km/s}} \qty(\frac{H}{17})^{-1}\nonumber\\
&\times \qty(\frac{\rho}{10^5\,\msun/\si{pc^3}})^{-1} \qty(\frac{a}{\SI{1}{mpc}})^{-1}
}
} 
From the scattering experiment data, we calculate these parameters in the following way (see Appendix~\ref{appendix:derivation} for the derivation):
{
\bsub\label{eq:HJK}
\barr
H &=& 2\pi\sigma \frac{(1+q)^2}{q} 
\left\langle p^2_\mathrm{max}v^2f(v)\frac{\delta E_\ast}{m_\ast} \right\rangle,\label{eq:H-numerical}\\
J &=& \frac{\pi\sigma}{H} 
\left\langle 
p^2_\mathrm{max}v^2f(v)\,u(v_\mathrm{ej}-5.5\sigma)
\right\rangle,\label{eq:J-numerical}\\
K &=& \frac{1-e^2}{2e}\qty(\frac{\left\langle p^2_\mathrm{max}v^2f(v)\delta L_{z,\ast}\right\rangle}{\sqrt{1-e^2}\left\langle p^2_\mathrm{max}v^2f(v)\delta E_\ast\right\rangle}-1),
\label{eq:K-numerical}\\
p^2_\mathrm{max} &=& 25\qty(1+0.4v^{-2}).
\earr
\esub
Here $\langle\dots\rangle$ means the average over all our simulations, $f(v)$ is the Maxwellian velocity distribution with dispersion $\sigma$, normalized so that $\int_0^\infty f(v)\dd{v}=1$:
\eq{
f(v) = \frac{v^2}{(2\pi)^{3/2}\sigma^3} e^{-v^2/2\sigma^2},
}
$u$ is the Heaviside step function,
}
$p_\mathrm{max}$ is the maximum value of the initial impact parameter (see Section~\ref{section:methods}) and $m_\ast$, $v$, $\delta E_\ast$ and $\delta L_{z,\ast}$ are the stellar mass, initial velocity and changes in energy and angular momentum, respectively. These equations also assume $a=v_0=1$.

Fig.~\ref{fig:HJ} (top) shows the values of $H$ and $J$ for different parameters of the binary. The values we got are in good agreement with \cite{Sesana2006}. We see that both of them are fairly independent of $q$ and $e$ and, following \cite{Sesana2006}, can be analytically approximated as
\bsub\label{eq:HJ}
\barr
H &=& A_H(1+a/a_{0,H})^{\gamma_H},\label{eq:H}\\
J &=& A_J(a/a_{0,J})^{\alpha_J} (1+(a/a_{0,J})^{\beta_J})^{\gamma_J}.\label{eq:J}
\earr 
\esub
The parameter values for these approximations are listed in Tables \ref{table:H} and \ref{table:J}. {The approximations are accurate within $5\%$.} We also consider both {maximally} corotating/counterrotating stellar nuclei ({ i.e. $\eta=1$ and $\eta=0$, respectively;} Fig.~\ref{fig:HJ}, bottom). $H$ is $\sim2$ times higher in the corotating case compared to the counterrotating one, and this difference persists for all values of $a/a_h$ (Fig.~\ref{fig:HJ}, bottom left)\footnote{This is in contrast to the equal-mass binary case studied in \cite{Rasskazov2017} where $H_\mathrm{corot} = H_\mathrm{counterrot}$ for $a\ll a_h$ and $H_\mathrm{corot} \approx -H_\mathrm{counterrot} < 0$ for $a\gg a_h$.}. As for the dependence of $J$ on rotation, it is more complex: $J_\mathrm{corot} \approx 0.5J_\mathrm{counterrot}$ for $a/a_h\ll0.3$ and $J_\mathrm{corot} \approx 2J_\mathrm{counterrot}$ for $a/a_h\gg0.3$ (Fig.~\ref{fig:HJ}, bottom right). 
We have also performed calculations to emulate a co-/counter-rotating stellar disk by only considering stars with $\theta<\pi/20$ and $\theta>19\pi/20$, respectively. In this case, the difference in co- and counter-rotating $H$ and $J$ is more pronounced: a factor of 3-5 instead of 2.

Fig.~\ref{fig:K} shows the values of $K$ for different parameters of the binary{; we found their dependence on $a/a_h$ to be too complex to be approximated by a simple analytical expression similar to Eqs.~\eqref{eq:HJ}.}
Even though they agree with \cite{Quinlan1996}, our values of $K$ differ significantly from those of \cite{Sesana2006}: they are systematically higher by up to a factor of 1.5. It turns out, there is a mistake in Sesana et al.'s procedure for the calculation of $K$: they calculate $K$ for a single-velocity stellar distribution \citep[][Eq. 12]{Sesana2006} and then average it over the Maxwellian distribution -- the correct way is instead to separately velocity-average the enumerator $(\dv*{e}{t})$ and the denominator ($\dv*{\ln a}{t}$) in the definition for $K$. For this reason, we're also 
{ showing} the $K$ values for mass ratios $q>10^{-2}$ which are ruled out for the case of an IMBH in our galactic center \citep{mer13}. 

{Curiously, somewhere between {$q=10^{-3}$ and $q=10^{-2}$} (the regime unexplored in any previous work) $K$ for non-rotating case switches from mostly positive to mostly negative
The reasons for that are unclear but that implies we shouldn't expect { the IMBH orbit}
to be eccentric.}

Fig. \ref{fig:K1} shows $K$ for co- and counterrotating stellar environments. {For $q\lesssim10^{-2}$, we have confirmed the previous findings that the binary eccentricity tends to decrease in a corotating environment and increase in a counterrotating one \citep{2011MNRAS.415L..35S,Rasskazov2017}, $|K|$ being about an order of magnitude larger compared to the nonrotating case and weakly dependent on $a/a_h$ (Fig. \ref{fig:K1}, top). However, for $q\lesssim10^{-3}$ the rotation seems to be less significant (Fig. \ref{fig:K1}, bottom).} An equal-mass corotating binary presents a special case as $K$ becomes infinite at $a/a_h\approx10$; it can be fitted as
\barr\label{eq:K(q=1)}
K = -26.6\,(10-a/a_h)^{-1.77}.
\earr
This happens because $H$ becomes negative for sufficiently soft equal-mass binaries, as was discovered in \cite{Rasskazov2017}. 

\begin{figure*}
	\centering
	\subfigure{\includegraphics[width=0.49\textwidth]{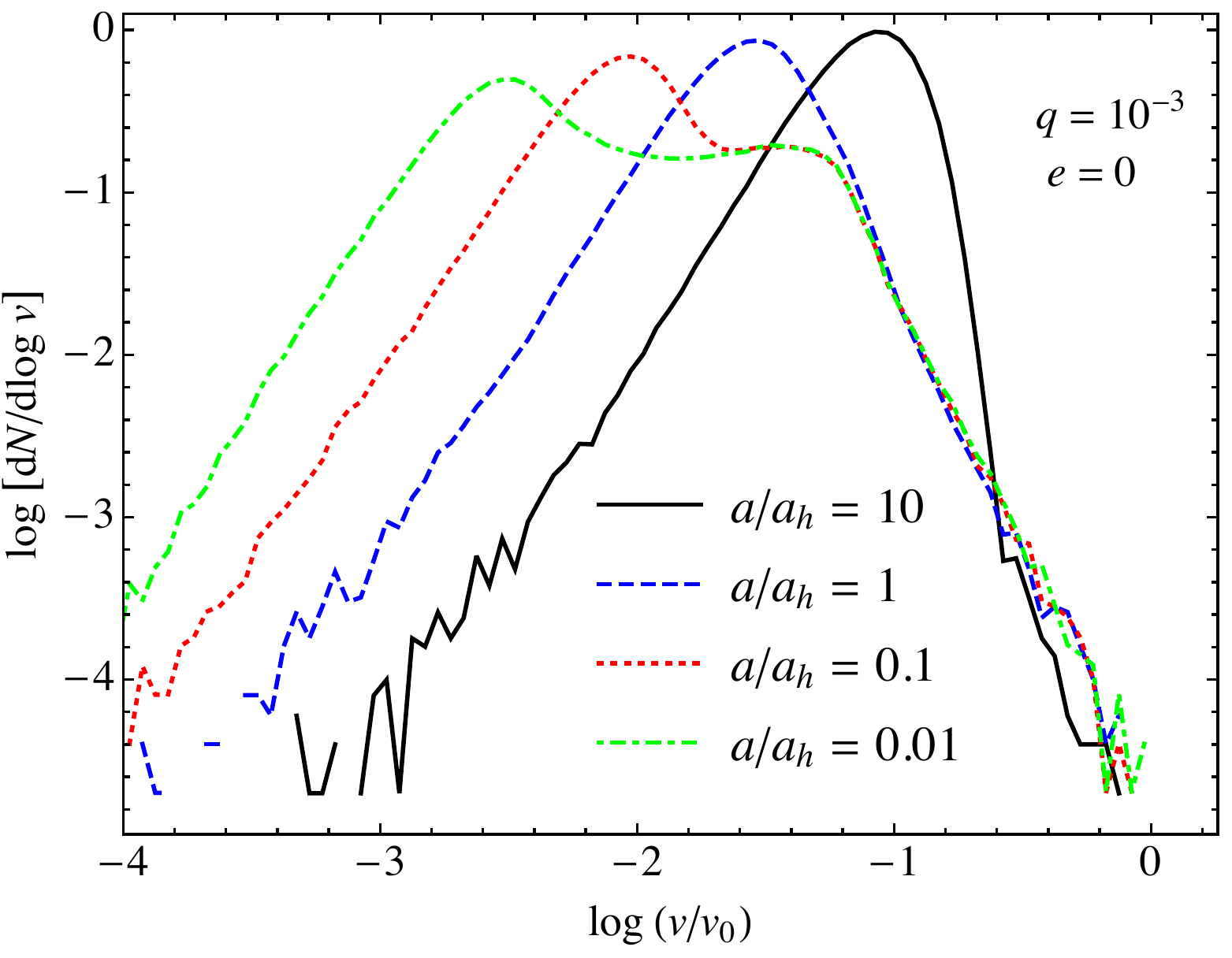}}
	\subfigure{\includegraphics[width=0.49\textwidth]{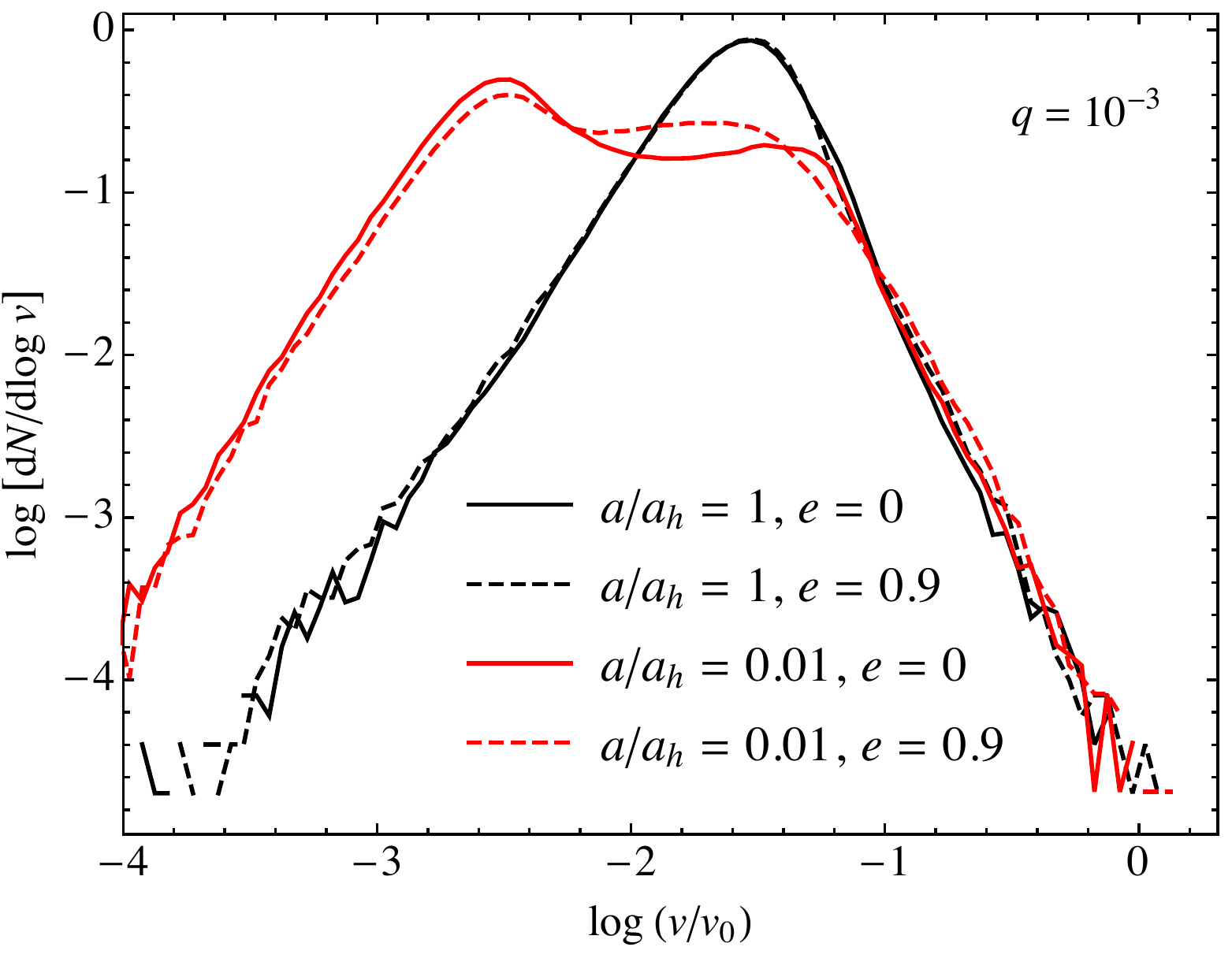}}
	\subfigure{\includegraphics[width=0.49\textwidth]{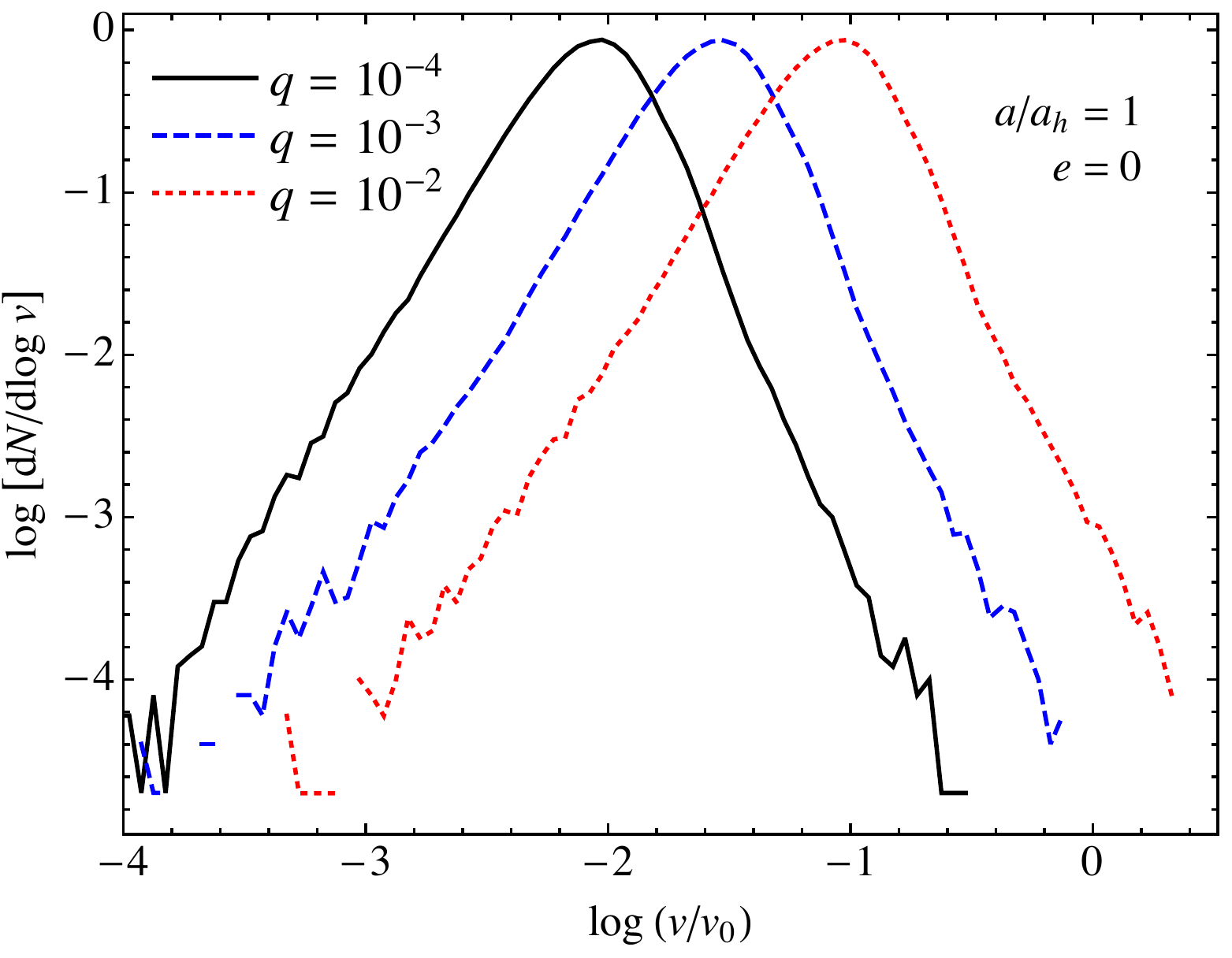}}
    \subfigure{\includegraphics[width=0.49\textwidth]{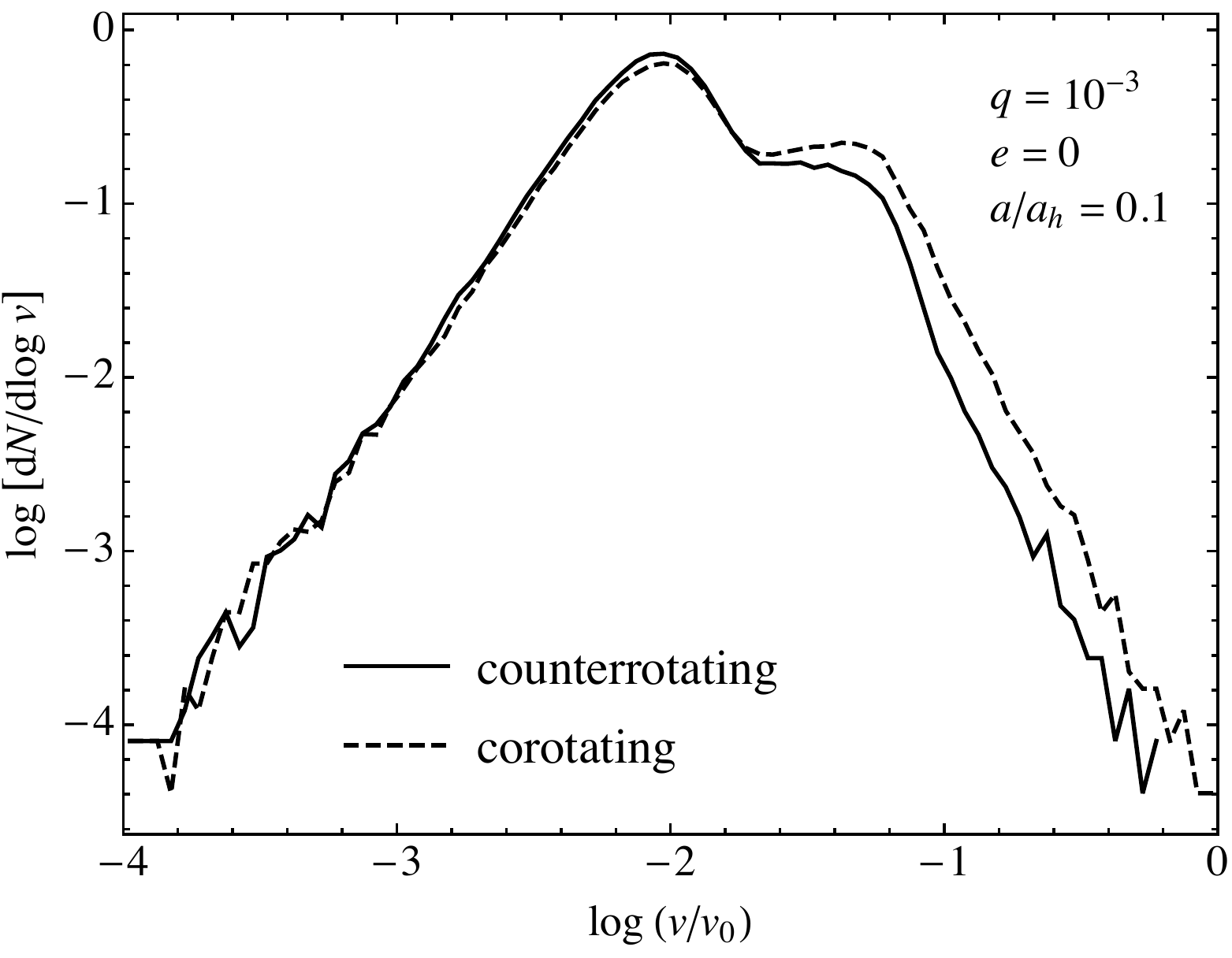}}
    \subfigure{\includegraphics[width=0.49\textwidth]{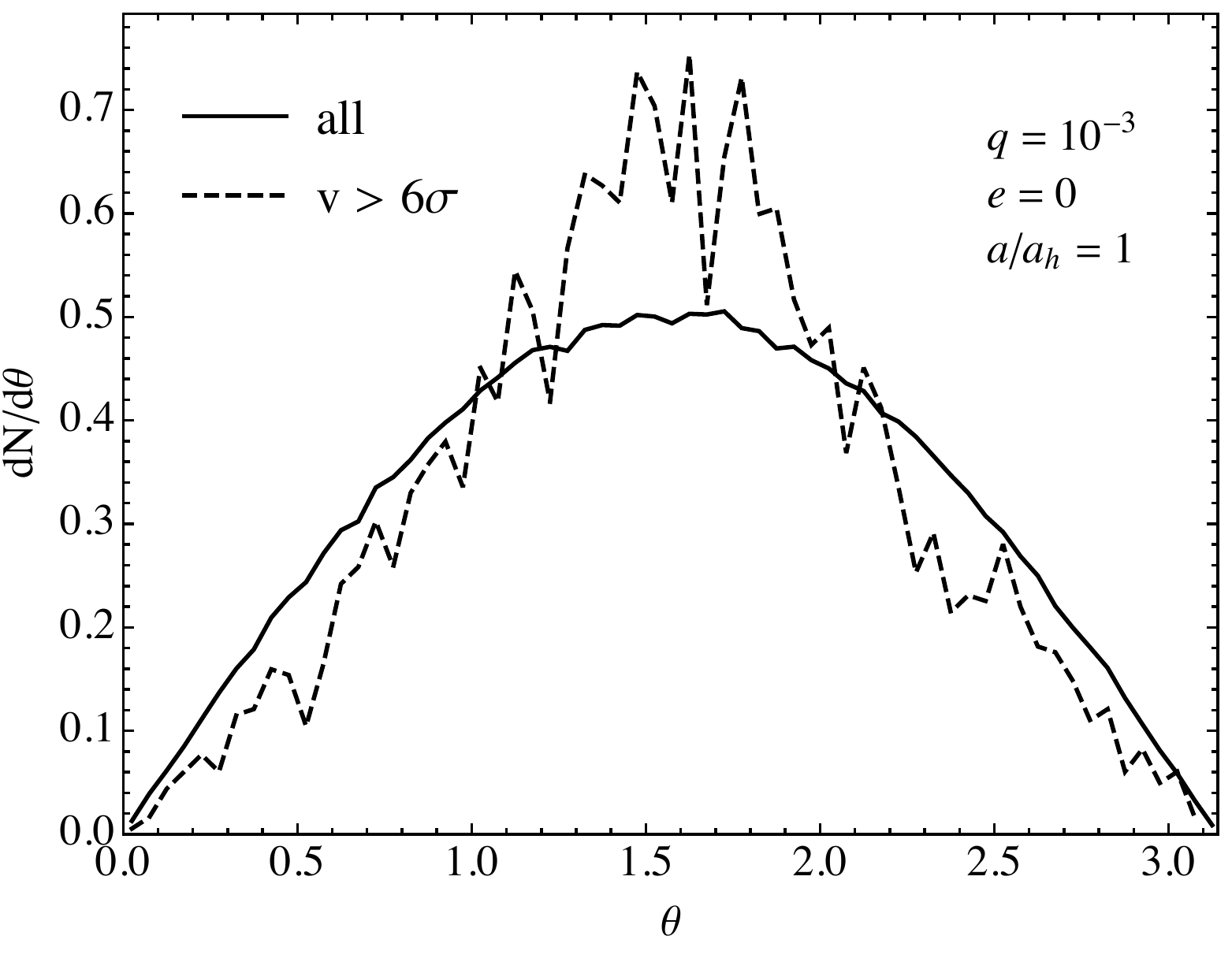}}
    \subfigure{\includegraphics[width=0.49\textwidth]{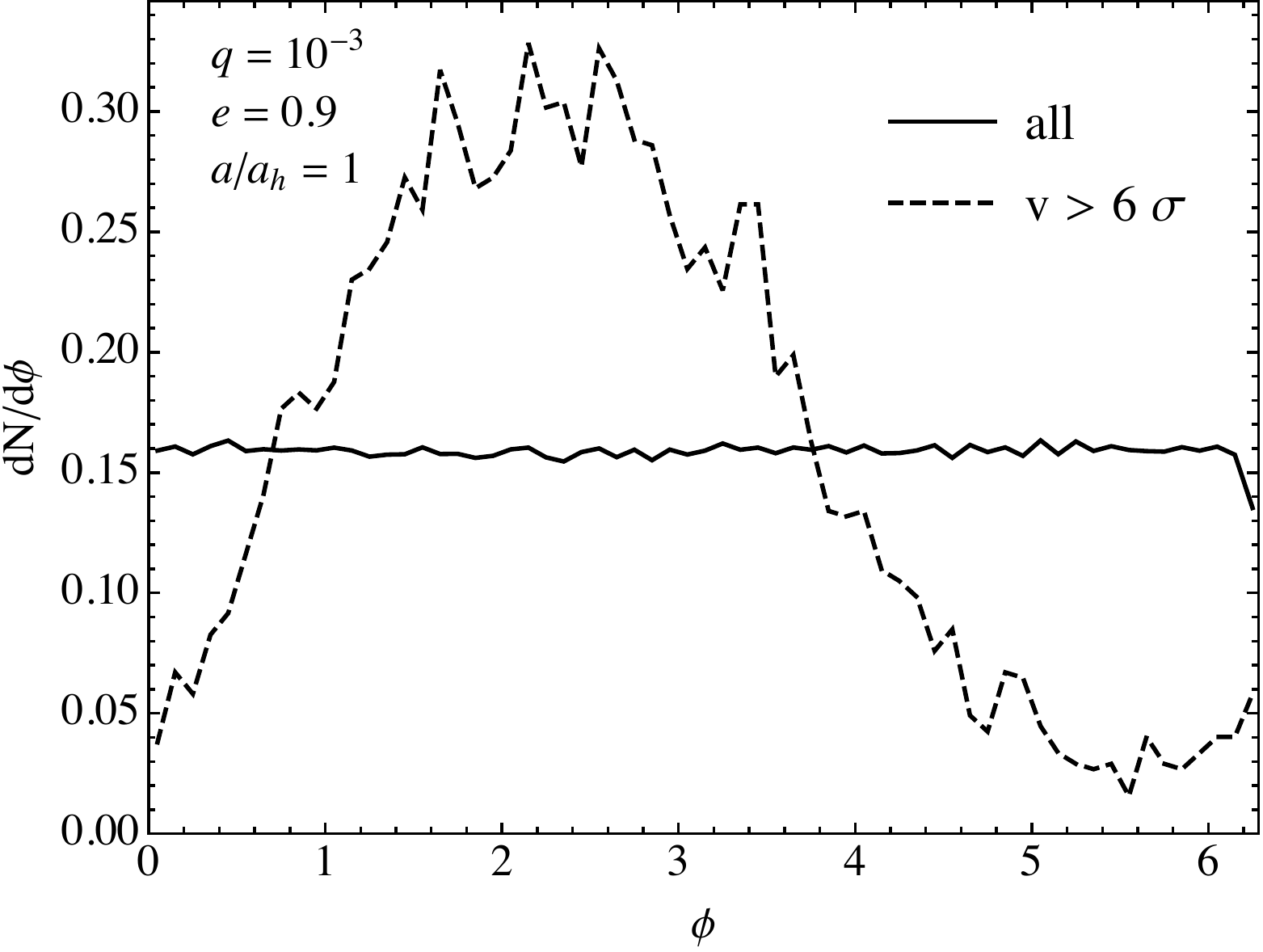}}
\caption{
Top and middle: the distribution of ejection velocities (in units of the binary orbital velocity $\qty(v_0 = \sqrt\frac{GM}{a})$) for different parameters of the binary. Bottom: distribution of the direction angles of the stellar escape velocity, where $\theta=0$ is the direction of the massive BH binary angular momentum and $\phi=0$ is the direction of the SBH orbital pericenter. All distributions are normalized such that the total number is equal to 1.
}
\label{fig:v}
\end{figure*}

Fig.~\ref{fig:v} (top and middle) shows the distribution of ejected stars' velocities for various binary parameters. These are in good agreement with \cite{Sesana2006}.\footnote{Note that the plots in the current paper show $\dv{N}{\log{v}} = v\dv{N}{v}$.} The maximum is due to the initial (Maxwellian) velocity distribution having a peak at $\sim \sigma = \frac{1}{2}v_0 \sqrt{\frac{a}{a_h}\frac{q}{1+q}}$. The component of this distribution which is relevant to the HVS problem is the high-velocity tail $v>\vmin$,
\barr
\vmin &=& v_0 \frac{\sqrt{2q}}{1+q} = 
2\sqrt{2}\sigma\sqrt\frac{a_h}{a} \nonumber\\
&\approx& 
\SI{280}{km/s} \sqrt\frac{a_h}{a} \frac{\sigma}{\SI{100}{km/s}}.
\earr
In this hypervelocity tail $\dv{N}{\log{v}} \sim v^\alpha$, $\alpha\approx-3$ irregardless of the binary parameters, including rotation. {Note also that the velocity distribution, including $\vmin$, doesn't depend on $q$ explicitly (only through $a_h$).} The absolute number of stars in the tail ejected per unit time is determined by $J$ and $H$ (the velocity limit of $5.5\sigma$ in the definition for $J$ falls within the tail unless $a\ll a_h$ when the massive BH binary is in the GW-dominated regime). We find that the harder the binary gets, the faster are the HVSs it generates. Combined with the steep decrease of $J(a/a_h)$, this implies that most HVSs will be ejected during the later stages of the binary lifetime.

The bottom row of Fig.~\ref{fig:v} shows the distribution of HVS velocity directions. All the ejected stars together are distributed almost spherically ($\propto\sin\theta$ and uniformly in $\phi$) as most stars interact with the binary very weakly. However, if we only consider the stars fast enough to escape from the GC (taking the escape velocity above $\sim6\sigma$), they are preferentially ejected in the direction of the massive BH binary orbital plane, as was also found in \citet{Sesana2006}. The distribution of $\phi$ for these fast stars is also not uniform, with a maximum around $\sim3\pi/4$.  Azimuthal anisotropy has also been observed by \citet{Sesana2006}, although they found that the HVSs are preferentially ejected in the direction of the secondary BH velocity at pericenter ($3\pi/2$); the reasons for this discrepancy remain unclear, but we return to this interesting discrepancy in Section~\ref{section:conclusions}.

\section{Hills Mechanism}
\label{section:hills}

\begin{figure}
\centering
\includegraphics[width=0.45\textwidth]{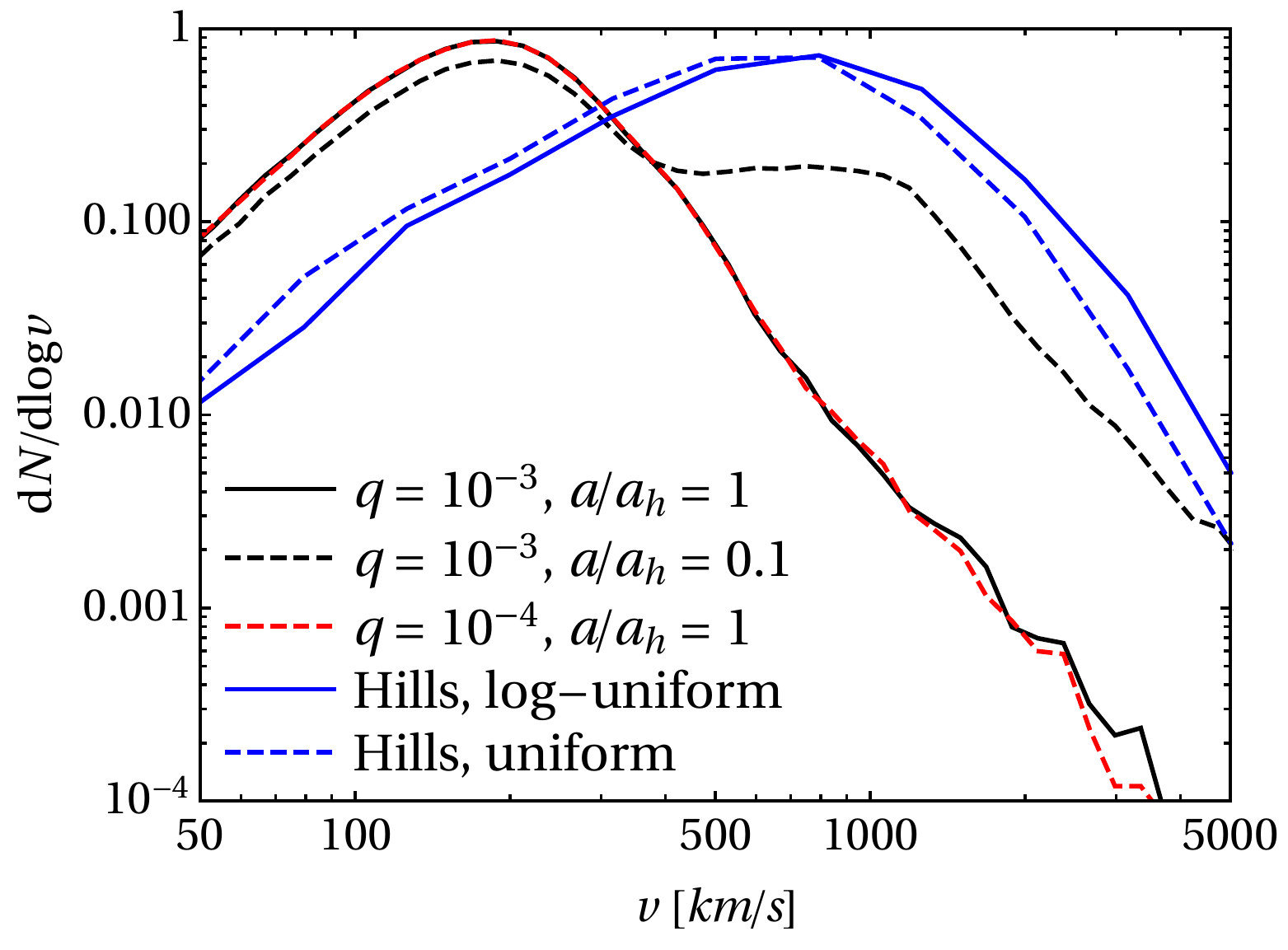}
\caption{Comparison between the expected ejection spectrum of the Hills mechanism and the SMBH-IMBH scenario. In the \citet{hills88} case, we consider both stars with masses $m_1=m_2=1$ M$_\odot$. The initial distribution of semi-major axes is taken to be log-uniform in the range $0.01$ AU-$1$ AU, while the eccentricity distribution is taken to be thermal. In the SMBH-IMBH scenario, we take $q=10^{-4}$-$10^{-3}$ and $a/a_h=0.1$-$1$, which falls within the allowed parameter space of the putative IMBH in the Galactic Center \citep{Merritt2002}. All distributions are normalized to unit area.}
\label{fig:comparhills}
\end{figure}

In this section, we briefly summarize the standard \citet{hills88} scenario. A binary of total mass $m_b$ and semi-major axis $a_b$ is tidally disrupted whenever it approaches a single SMBH of mass $M_{\rm SMBH}$ within a distance of order the binary star tidal radius
\begin{equation}
r_t \approx 10~{\rm AU} \left(\frac{a_{\rm b}}{0.1 \rm AU}\right) \left(\frac{M_{\rm SMBH}}{m_{\rm b}/ 4 M_{\odot}}\right)^{1/3},
\label{eqn:rt}
\end{equation}
where here and the following we set $M_{\rm SMBH} = 4 \times 10^{6} M_{\odot}$ when a numerical value is needed.
The distance of closest approach can be computed using angular momentum conservation. If a binary star starts from an initial distance $D$ with transverse speed $v$, the distance of closest approach $r_{\rm min}$ will be
\begin{equation}
v\,D=\left(\frac{GM_{\rm SMBH}}{r_{\rm min}}\right)^{1/2}r_{\rm min}\ .
\label{eqn:rmin}
\end{equation}
In the case $r_{\rm min} \lesssim r_t$, the binary star is tidally separated and the former companion is dissociated. There are three possible outcomes for the tidal breakup of a binary\footnote{Further possibilities for  binary crossing $r_{\rm t}$, not considered in this paper, are that either of the stars or both cross their tidal {\it disruption} radius $r_{\rm d}/r_{\rm t} \sim (R_*/a_{\rm b}) \ll 1$, (where $R_*$ is the stellar radius) and get tidally disrupted, or that the binary members merge \cite[e.g.][]{mandel_levi15,bonnerot18}.}: (i) production of an HVS and an S-star {(a star orbiting around the SMBH)}, (ii) production of 2 S-stars, (iii) capture of the whole binary. From energy conservation, these two latter can occur only for binaries whose center of mass is on a bound --elliptic-- obit, while in the former case the approaching binary should be close to a parabolic trajectory\footnote{During the tidal breakup, the two stars gain the opposite amounts of energy, and the binding energy of the stellar binary can be neglected. Therefore, the outcome where one star is ejected, and the other one stays bound, requires the binary to have zero orbital energy, i.e. being on a parabolic orbit.}. In the case of a triple \citep{perets2009,gins2011,fgu18} or quadruple star \citep{fragio18}, even more outcomes are possible.

For a binary centre of mass approaching on a nearly parabolic orbit, the ejected star has a velocity at infinity with respect to the black hole of \citep{hills88,brm06,sari10,kobay12}
\begin{eqnarray}
v_{\rm ej}& =&K_{\rm g} \sqrt{\frac{2 G m_{\rm c}}{a_{\rm b}}}  \left(\frac{M_{\rm SMBH}}{m_{\rm b}}\right)^{1/6} \nonumber\\
& \approx & 1500 ~{\rm \frac{km}{s}} ~ \sqrt{\frac{m_{\rm c}/M_{\odot}}{a_{\rm b}/0.1 \rm AU}}  \left(\frac{M_{\rm SMBH}}{m_{\rm b}/ 2 M_{\odot}}\right)^{1/6},
\label{eqn:vej}
\end{eqnarray}
where $m_{\rm c}$ is the mass of companion star, that remained bound to the SMBH and $K_{\rm g}$ is a numerical coefficient of order unity that only depends on the specific geometry of the encounter (e.g. binary phase and binary pericenter, see figure 10 in \citet{sari10}). For a given binary, 
\citet{brm06} and \citet{fgu18} numerically showed that the overall distribution is well approximated by a Gaussian distribution with mean $v_{\rm ej}$ and dispersion $\sigma\sim 0.2$-$0.3 \ v_{ej}$. The broadening of the distribution around the Hills peak $v_{\rm ej}$ is due to averaging over the random orientations of the initial binary phases, orientations and eccentricities.
We also remark that for a parabolic orbit the ejection probability is exactly 50\% for the primary, while the primary gets preferentially captured (ejected) for eleptical (hyperbolic) approaching trajectories \citep{sari10,kobay12}. The velocity distribution for a population of binary with mass and semimajor axis distributions can then be analytically and accurately approximated using Eq. (\ref{eqn:vej}) with $K_{\rm g}=1$ \citep{ross14}.

To compare the expected ejection velocity distribution obtained in the SMBH-IMBH case to the Hills mechanism case, we run $1$ million 3-body simulations of the close encounter of a binary star and an SMBH.  Also in this case, we use the \textsc{archain} algorithm to accurately integrate the motions of our systems \citep{Mikkola2008}. We set the initial conditions of our 3-body simulations following the prescriptions of \citet{gl06,gl07} and \citet{fgu18}. We consider equal-mass binaries, and sample the initial distribution of semi-major axes ($a_b$) from a uniform or log-uniform distribution, in the range $0.01$ AU-$1$ AU \citep{frasar18}. We also assume a thermal distribution of binary eccentricities ($e_b$), and check that the sampled binaries satisfy $a_b(1-e_b)>R_*$ ($R_*$ is the stellar radius). We then randomly sample the pericenter distance ($r_p$) of the binary according to a probability $\propto r_p$, in the range $0-2R_t$, where $R_t=r_t(a_b=1\ \mathrm{AU})$, i.e. twice the maximum tidal radius of the softest binary in our semi-major axis distribution. We note that this choice for the distribution of $r_{\rm p}$ implies that wider binaries are preferentially separated because their tidal radius $r_{\rm t} \propto a_{\rm b}$ is larger, enhancing lower velocity ejections (see eq.\ref{eqn:vej}). We note, however, that this is not the only possible way to distribute $r_{\rm p}$, as binaries that diffuse in angular momentum space by two-body encounters would be instead mostly separated at $r_{\rm p} \sim r_{\rm t}$, implying a disruption probability independent of $a_{\rm b}$.\footnote{Factually, we are assuming that the refilling of the loss cone with binary orbits occurs in the "full" regime \citep{shapiro77,ross14}.} Given $a_b$ and $e_b$ of a specific sampled binary, we integrate the 3-body system if $r_p<2r_t$, otherwise we reject the binary parameters.

Figure \ref{fig:comparhills} shows a comparison between the expected ejection spectrum of the Hills mechanism and the SMBH-IMBH scenario. In the \citet{hills88} case, we consider both stars to have masses $m_1=m_2=1$ M$_\odot$. In the SMBH-IMBH scenario, we take $q=10^{-4}$-$10^{-3}$ and $a/a_h=0.1$-$1$, within the allowed parameter space of the putative IMBH in the GC \citep{Merritt2002}.  In the case $a/a_h=1$, the SMBH-IMBH scenario predicts much smaller final velocities, with a distribution peaked at $\sim 200\,\kms$, while the Hills scenario typically produces higher velocities peaked at $\sim 800\, \kms$. However, for a smaller SMBH-IMBH semi-major axis $a/a_h=0.1$ and stellar velocities $v\gtrsim\SI{500}{km/s}$ \citep[about the minimum velocity for a star to become a HVS,][]{Sesana2006}, the SMBH-IMBH scenario distribution looks similar to  the Hills scenario {(up to normalization): approximately constant under $\SI{1000}{km/s}$ and quickly falling off ($\propto v^{-3}$) above that. However, the actual IMBH scenario distribution would be a convolution of distributions at different single values of $a/a_h$ that takes into account the evolution of $a$:
\eq{
\dv{N}{v} = \frac{1}{T} \int_0^T \dd{t} \dv{a}{t} \frac{\dd N}{\dd{v}\dd{a}}
}
where $T$ is the SMBH-IMBH binary lifetime and $\dd{N}/\dd{v}\dd{a}$ is the number of stars ejected as the binary shrinks from $a+\dd{a}$ to $a$. Calculating that distribution and comparing it with observations will be the subject of our next paper. 
}

In the Hills scenario, we note that we do not find significant differences between adopting a log-uniform and uniform distribution in the binary semi-major axis. {The main reason is that, while in the log-uniform case a larger number of tight binaries is sampled, these binaries typically are not disrupted by the SMBH (and only experience a flyby) because of a small tidal radius. Nevertheless, the high-velocity tail of the distribution depends on the assumed period distribution of binaries. Figure~\ref{fig:comparhills} shows that the log-uniform case has typically larger velocities in the high-velocity tail of the distribution.} We have also run models with $m_1=m_2=3$ M$_\odot$. Also in this case, we do not find any significant difference with the case $m_1=m_2=1$ M$_\odot$. This is because, while a larger mass implies typically a larger ejection velocity (Eq.~\ref{eqn:vej}),  the sampled initial binary semi-major axes are typically larger (to satisfy $a_b(1-e_b)>R_*$ ). 

\section{Conclusions}
\label{section:conclusions}

We have performed a series of 3-body scattering experiments of an initially unbound star getting ejected by a potential SMBH-IMBH binary in our galactic center. The IMBH mass was assumed to be in the range $q = 10^{-4}-10^{-2}$ of the SMBH mass. The new features compared to previous papers are 1) scattering experiments for $q \leq 10^{-3}$ and 2) using the ARCHAIN numerical code for the scattering experiments.

We found the distribution of HVS velocities to be a power-law $(\dv*{N}{v})\propto v^{-4}$ regardless of the binary or stellar parameters. The HVS velocity directions are concentrated around the binary orbital plane and, in the case of an eccentric binary, they have a preferred direction in the binary plane. 
We have also calculated the stellar ejection rate $J$ as well as the evolution rates of the binary semimajor axis $H$ and eccentricity $K$. We found $J$ and $H$ to be fairly independent of $q$. $K$, however, changes at the lowest values of $q$ we have probed ($10^{-4}-10^{-3}$), becoming negative at most values of $a$ and $e$. Therefore, we should probably expect the IMBH orbit to be circular. This result is unexpected and deserves further investigation as $K$ shows quite a consistent behavior at $q=1-10^{-2}$. Adding rotation of the stellar nucleus does not affect that conclusion (see below).

{
When calculating the diffusion coefficients $H$, $J$ and $K$ we have ignored the depletion of the loss cone -- i.e., the region in phase space where the stars can approach the BH binary closer than $\lesssim a$. We effectively consider the loss cone to be always full; its depletion would mean the slow-down (or even a complete stall) of BH binary orbital evolution, i.e. the decrease in $H$ and $J$. As for $K$, it should not change as both $\dv*{e}{t}$ and $\dv*{a}{t}$ are proportional to the number of stars interacting with the binary.
The loss cone is repopulated through 2-body relaxation and stellar angular momentum drift in nonspherical galaxies \citep[][Chapter 4]{mer13}. Previous studies to date have focused on the evolution of supermassive binaries. It was shown that in spherical galaxies the relaxation is not strong enough to keep the loss cone full, which causes the BH binary evolution to stall \citep[``the final parsec problem'',][]{Milosavljevic2003,Vasiliev2014}. However, even a small degree of triaxiality is enough to overcome this problem and keep the binary shrinking \citep{Vasiliev2015}; \citet{Khan2013} claim this to be true for axisymmetric galaxies as well. 
As shown in \citet{Vasiliev2015}, it is straight-forward to incorporate the loss-cone depletion in galaxies with different geometries using a coefficient depending on $a/a_h$ \citep[][Eq. 21]{Vasiliev2015}:
\eq{
H = \mu \qty(\frac{a}{a_h})^\nu H_{\rm Full\,LC},
}
where $\mu\lesssim1$, $\nu=0.3\dots0.6$ for triaxial galaxies and $\nu\approx0$, $\mu\ll1$ for axisymmetric and spherical galaxies; the analogous relation should hold for $J$ as well. This key approximation, combined with $H_{\rm Full\,LC}$ and $J_{\rm Full\,LC}$, derived in this paper, will be used in our next paper where we compute the time-integrated HVS velocity distributions and compare to the observed data.
}

We found two discrepancies with the results of \citet{Sesana2006}. First, our values of $K$ are up to 1.5 times higher due to \citet{Sesana2006} using an incorrect procedure to calculate $K$. Also, our simulations show a different distribution for $\phi$ -- the direction of the ejection velocity in the plane of the binary -- which is non-uniform for eccentric binaries. Its peak is not at $3\pi/2$ as in \cite{Sesana2006} (the direction of the IMBH velocity at its pericenter) but rather at $\lesssim\pi$. We could not determine the source for this discrepancy. However, the difference would be hard to test observationally unless we have independent constraints on the IMBH pericenter (and \citet{Sesana2006} do not provide the exact distribution of $\phi$ in their paper). { In addition to that, we haven't accounted for the general-relativistic pericenter precession which has a characteristic timescale similar to the hardening timescale \citep{Rasskazov2017} and therefore can make the overall distribution of $\phi$ uniform (i.e. make the ejection directions distribution axisymmetric).}

Another new result of this paper is simulations of a corotating/counterrotating stellar nucleus where only the stars with positive/negative $L_z$ were selected. We find the following:
\begin{itemize}
\item Corotation/counterrotation decreases/increases the hardening rate by $\sim50\%$. 
\item Corotation decreases/increases the stellar ejection rate for binary semimajor axes below/above $\sim0.5a_h$ (the opposite for counterrotation). 
\item {For $q\gtrsim10^{-2}$, eccentricity tends to decrease/increase in corotating/counterrotating systems, and 
the absolute value of $K$ is much higher compared to the nonrotating case. For $q\lesssim10^{-3}$, however, eccentricity is { decreasing  regardless of rotation.}
}
\end{itemize}

{However, our model only takes into account the ejection of unbound stars. The stars that were initially bound to the SMBH can also interact with the IMBH and change its orbital elements \citep{levin2006,baumg2006,Matsubayashi2007}. Also, some of the initially unbound stars get captured on very weakly bound (almost parabolic) orbits for a very long time, and while we ignore their effect, they can affect the binary orbit as well.}

Finally, we have also compared the SMBH-IMBH ejection scenario to the standard Hills mechanism. We find that { the ejection velocity distributions can be similar given $a/a_h\sim0.1$}
(see Figure \ref{fig:comparhills}). 
{ The distributions of ejection directions, however, would be different: e.g. given a spherically symmetric source of incoming stars, the Hills mechanism produces a spherically symmetric distribution of ejected stars, while the IMBH mechanism preferentially ejects stars in the IMBH orbital plane.}
Also, the high velocity tail in the Hills mechanism crucially depends on the assumed star binary mass-ratio and separation distributions, which we are not widely exploring here. In particular, softer high velocity tails can be obtained with star binary distributions consistent with current observations \citep{rossi2017}. 

These results will be used in our next paper to construct samples of ejected stars, where we will take into account the moment of ejection, orbit in the Milky Way potential and also the SMBH-IMBH binary evolution for the case of IMBH slingshot ejections. These samples will then be compared to the observed HVS distributions (using mixture models to combine all of the HVS production scenarios described in this paper) to test the hypothesis that they might have been generated by one of the two mechanisms we consider. 

\section*{Acknowledgements}

This research was started at the NYC Gaia DR2 Workshop at the Center for Computational Astrophysics of the Flatiron Institute in 2018 April. GF is supported by the Foreign Postdoctoral Fellowship Program of the Israel Academy of Sciences and Humanities. GF also acknowledges support from an Arskin postdoctoral fellowship and Lady Davis Fellowship Trust at the Hebrew University of Jerusalem. 
AR is supported by the European Research Council (ERC) under the European Union's Horizon 2020 research and innovation program ERC-2014-STG under grant agreement  No 638435 (GalNUC).

\appendix
\section{Calculation of $H$, $J$ and $K$}
\label{appendix:derivation}

By definition,
\barr
H = -\frac{\sigma}{G\rho a^2} \dv{a}{t} = -\frac{2\sigma}{\qty(GM)^2\rho} \frac{(1+q)^2}{q} \dv{E}{t}
\earr
where $E$ is the binary orbital energy. Assuming an infinite uniform distribution of stars\footnote{which is a valid approximation if we assume $\rho$ and $\sigma$ to be the density and velocity dispersion at the influence radius, as discussed in \citet[][Section 3.2]{Rasskazov2017}} \citep[cf.][Eq. 16a]{Merritt2002},
\barr\label{eq:dEdt}
\dv{E}{t} = \int_0^\infty \dd{v} \int_0^{p_\mathrm{max}(v)} 2\pi p\dd{p} v n f(v)
\left\langle\delta E\right\rangle_{\theta,\phi} 
= -\int_0^\infty \dd{v} \int_0^{p_\mathrm{max}(v)} 2\pi p\dd{p} v n f(v) 
\frac{\left\langle\delta E_\ast\right\rangle_{\theta,\phi}}{m_\ast}
\earr
where $\delta E$ and $\delta E_\ast$ are changes in the binary energy and stellar energy, respectively, for a single scattering event, $f(v)$ is the initial stellar velocity distribution $\qty(\int_0^\infty f(v)\dd{v}=1)$, $n=\rho/m_\ast$ is the stellar number density, and $\langle\dots\rangle_{\theta,\phi}$ means averaging over all angles. 
{
As $v$ and $p$ in our simulations are distributed uniformly in $\log{v}$ and $p^2$, $H$ can be calculated using the simulation results in the following way:
\eq{
H = 2\sigma\frac{(1+q)^2}{q} \int_0^\infty \frac{\dd{v}}{v} \int_0^{p_\mathrm{max}(v)} \frac{2 p\dd{p}}{p_\mathrm{max}^2} \pi p_\mathrm{max}^2 v^2 f(v) 
\frac{\left\langle\delta E_\ast\right\rangle_{\theta,\phi}}{m_\ast} 
= 2\pi\sigma\frac{(1+q)^2}{q} \left\langle p_\mathrm{max}^2 v^2 f(v) \, \frac{\delta E_\ast}{m_\ast} \right\rangle \ln\frac{v_{\rm max}}{v_{\rm min}}
}
where we have used the $GM=a=1$ units and 
\eq{
\langle F \rangle \equiv 
\frac {\int_0^\infty \frac{\dd{v}}{v} \int_0^{p_\mathrm{max}(v)} \frac{2p\dd{p}}{p_\mathrm{max}^2} F(p,v)} {\int_0^\infty \frac{\dd{v}}{v} \int_0^{p_\mathrm{max}(v)} \frac{2p\dd{p}}{p_\mathrm{max}^2}} =
\frac {\int_0^\infty \frac{\dd{v}}{v} \int_0^{p_\mathrm{max}(v)} \frac{2p\dd{p}}{p_\mathrm{max}^2} F(p,v)} {\ln\frac{v_{\rm max}}{v_{\rm min}}}
}
means the average over our set of simulations (i.e. over $v$, $p$ and all angles). 
}
{
In a similar manner, we can derive the expressions for $J$ and $K$:
\bsub
\barr
J &=& \frac{1}{M} \frac{\dv*{M_\mathrm{ej}}{t}}{\dv*{(1/a)}{t}} \frac{1}{a} 
= \frac{\sigma}{HGMa\rho} \dv{M_\mathrm{ej}}{t} 
= \frac{\sigma}{H} \int_0^\infty \dd{v} \int_0^{p_\mathrm{max}(v)} 2\pi p\dd{p} v f(v) \, u(v_\mathrm{ej}-5.5\sigma)
\nonumber\\
&=& \frac{\pi\sigma}{H} \left\langle v^2f(v)\,u(v_\mathrm{ej}-5.5\sigma) \right\rangle \ln\frac{v_{\rm max}}{v_{\rm min}}
\earr
\esub
where $u$ is the Heaviside step function and $v_\mathrm{ej}$ is the ejection velocity, and
\bsub
\barr
K &=& -\dv{e}{t}\qty(\dv{a}{t})^{-1}a = -\frac{1-e^2}{e}\frac{m_\ast}{M} \qty(\frac{M}{\mu L_z}\dv{L_{z,\ast}}{t} - \frac{a}{G\mu}\dv{E_{\ast}}{t}) 
\qty(-\frac{2a^2m_\ast}{GM\mu})^{-1} a \nonumber\\
&=& \frac{1-e^2}{2e} \qty(\frac{\dv*{L_{z,\ast}}{t}}{\sqrt{1-e^2}\dv*{E_{\ast}}{t}}-1)
= \frac{1-e^2}{2e} 
\qty(\frac{\left\langle p_\mathrm{max}^2 v^2 f(v) \,\delta L_{z,\ast}\right\rangle}
{\sqrt{1-e^2}\left\langle p_\mathrm{max}^2 v^2 f(v) \,\delta E_\ast\right\rangle}-1),\\
\mu &\equiv& \frac{M_1M_2}{M_1+M_2}.
\earr
\esub
}

\bibliographystyle{yahapj.bst}
\bibliography{hvs}

\end{document}